\documentclass[fa4paper,superscriptaddress,twocolumn,11pt,accepted=2018-08-05]{quantumarticle}

\usepackage[utf8]{inputenc}
\usepackage[english]{babel}
\usepackage[T1]{fontenc}
\usepackage{amsmath}
\usepackage{hyperref}

\usepackage{tikz}
\usepackage{lipsum}
\usepackage{amssymb}
\usepackage{graphicx}
\usepackage{graphics}
\usepackage{amsmath}
\usepackage{amsthm}
\usepackage{color}
\usepackage{dsfont}

\usepackage{mathtools}
\def\ra{\rangle}

\newtheorem{theorem}{Theorem}

\newtheorem{cor}[theorem]{Corollary}

\newcommand{\bea}{\begin{eqnarray}}
\newcommand{\eea}{\end{eqnarray}}
\newcommand{\be}{\begin{equation}}
\newcommand{\ee}{\end{equation}}
\newcommand{\ba}{\begin{equation}\begin{aligned}}
\newcommand{\ea}{\end{aligned}\end{equation}}

\newtheorem{definition}{Definition}
\theoremstyle{remark}
\newtheorem{remark}{Remark}

\def\be{\begin{equation}}
\def\ee{\end{equation}}

\newcommand{\mF}{\mathcal{F}}

\newcommand{\mH}{\mathcal{H}}

\newcommand{\mM}{\mathcal{M}}
\newcommand{\mN}{\mathcal{N}}
\newcommand{\mT}{\mathcal{T}}

\newcommand{\mW}{\mathcal{W}}
\newcommand{\mK}{\mathcal{K}}

\newcommand{\spa}{\text{span}}

\newcommand{\mS}{\mathcal{S}}

\newcommand{\lr}{\rangle\langle}

\newcommand{\tr}{{\rm Tr}}

\newcommand{\mbb}[1]{\mathbb{#1}}

\newenvironment{ruledtabular}{}{}

\def\>{\rangle}
\def\<{\langle}

\def\supp{ \mathrm{supp}}

\begin{document}

	\title{Monogamy of entanglement without inequalities}
	\date{\today}

	\author{Gilad Gour}
	\email{giladgour@gmail.com}
	\affiliation{ Department of Mathematics and Statistics and Institute for Quantum Science and Technology,
		University of Calgary, Calgary, Alberta T2N 1N4, Canada}
	
	\author{Yu Guo}
	\email{guoyu3@aliyun.com}
	\affiliation{Institute of Quantum Information Science, Shanxi Datong University, Datong, Shanxi 037009, China}

	\begin{abstract}
		
		We provide a \emph{fine-grained} definition of monogamous measure of entanglement that does not invoke any 
		particular monogamy relation. Our definition is given in terms of an equality, as opposed to inequality, 
		that we call the ``disentangling condition". We relate our definition to the more traditional one, by 
		showing that it generates standard monogamy relations. We then show that all quantum Markov states 
		satisfy the disentangling condition for any entanglement monotone. In addition, we demonstrate that 
		entanglement monotones that are given in terms of a convex roof extension are monogamous if they 
		are monogamous on pure states, and show that for any quantum state that satisfies the disentangling 
		condition, its entanglement of formation equals the entanglement of assistance. We characterize all 
		bipartite mixed states with this property, and use it to show that the G-concurrence is monogamous.
In the case of two qubits, we show that the equality between entanglement of formation and assistance holds 
if and only if the state is a rank 2 bipartite state that can be expressed as the marginal of a pure 3-qubit state in the $W$ class.
		
	\end{abstract}

	\maketitle

	Monogamy of entanglement is one of the non-intuitive phenomena of quantum physics that distinguish it 
	from classical physics. Classically, three random bits can be maximally correlated. For example, 
	three coins can be prepared in a state in which with 50\% chance all three coins show ``head", and 
	with the other 50\% chance they all show ``tail". In such a preparation, any two coins are maximally 
	correlated. In contrast with the classical world, it is not possible to prepare three qubits $A,B,C$ 
	in a way that any two qubits are maximally entangled~\cite{Coffman}. In fact, if qubit $A$ is maximally 
	entangled with qubit $B$, then it must be uncorrelated (not even classically) with qubit $C$. This 
	phenomenon of monogamy of entanglement was first quantified in a seminal paper by
Coffman, Kundu, and Wootters (CKW)~\cite{Coffman} for three qubits, and later on studied intensively 
in more general settings~\cite{Horodecki2009,Koashi,
		Gour2005pra,Osborne,
		Gour,Ouyongcheng,
		Ouyongcheng2007pra2,Hiroshima2007prl,Adesso,Kim,
		Kim2009,
		Kim2010jpa,Renxijun,
		Cornelio,Streltsov,
		Cornelio2013pra,
		Liusiyuan,
		Oliveira2014pra,Zhuxuena2014pra,Bai,Regula2014prl,
		Salini,Choi,Luo,Hehuan,Eltschka,Kumar2015,Zhuxuena2015pra,
		Kumar,Lancien,Lami,Song,Regula,Luo2016pra,Jung,
		Chengshuming,Allen,Liqiting,Camalet}.

	Qualitatively, monogamy of entanglement measures the shareability of 
	entanglement in a composite quantum system, i.e., the more two
	subsystems are entangled the less this pair has entanglement with the rest of the system. 	
	This feature of entanglement has found potential applications in many quantum information
	tasks and other areas of physics, such as quantum key
	distribution~\cite{Terhal,Pawlowski,Gisin}, classification of quantum
	states~\cite{Dur,
		Giorgi,Prabhu}, condensed-matter physics~\cite{Ma,Brandao,Garcia},
	frustrated spin systems~\cite{Rao}, statistical physics~\cite{Bennett},
	and even black-hole physics~\cite{Susskind,Lloyd}.

	A monogamy relation is quantitatively displayed as an inequality of the following form:
	\be\label{basic}
	E(A|BC)\geq E(A|B)+E(A|C)\;,
	\ee
	where $E$ is a measure of bipartite entanglement and $A,B,C$ are three subsystems 
	of a composite quantum system. It states that the sum of the entanglement 
	between $A$ and $B$, and between $A$ and $C$, can not exceed the entanglement 
	between $A$ and the joint system $BC$. While not all measures of entanglement 
	satisfy this relation, some do. Consequently, any measure of entanglement that 
	does satisfy~\eqref{basic} was called in literature \emph{monogamous}.

	Here we argue that this definition of monogamous measure of entanglement captures 
	only partially the property that entanglement is monogamous. This is evident 
	from the fact that many important measures of entanglement do not satisfy 
	the relation~\eqref{basic}. Some of these measures are not even additive 
	under tensor product~\cite{Shor2001prl,Shor2004cmp,Vollbrecht} (in fact, 
	some measures~\cite{gour2} are multiplicative under tensor product). 
	Therefore, the summation in the RHS of~\eqref{basic} is clearly only 
	a convenient choice and not a necessity. For example, it is well known 
	that if $E$ does not satisfy this relation, it is still possible to find 
	a positive exponent $\alpha\in\mbb{R}_{+}$, such that $E^\alpha$ satisfies 
	the relation. This was already apparent in the seminal work of~\cite{Coffman} 
	in which $E$ was taken to be the square of the concurrence and not the concurrence 
	itself.	More recently, it was shown that many other measures of entanglement 
	satisfy the monogamy relation~\eqref{basic} if $E$ is replaced by $E^\alpha$ 
	for some $\alpha>1$~\cite{Kumar,Zhuxuena2014pra,Zhuxuena2015pra,Osborne,Kim2009,Luo,Bai,Ouyongcheng2007pra2,Choi}.

One attempt to address these issues with the current definition of a monogamous 
measure of entanglement, is to replace the relation~(\ref{basic}) with a family of monogamy relations of the form,
$
E(A|BC)\geq f\Big(E(A|B),E(A|C)\Big),
$
where $f$ is some function of two variables that satisfy certain conditions~\cite{Lancien}. 
However, such an approach is somewhat artificial in the sense that the monogamy relations are not derived from more basic principles.

In this paper we take another approach to monogamy of entanglement which is more ``fine-grained" in nature, 
and avoid the introduction of such a function $f$. Therefore, our definition of monogamous measure 
of entanglement (see Definition~1 below) does not involve a monogamy relation, but instead a condition 
on the measure of entanglement that we call \emph{the disentangling condition} (following the terminology of~\cite{Hehuan}).
Yet, we show that our definition is consistent with the more traditional notion of monogamy of entanglement, 
by showing that $E$ is monogamous according to our definition if and only if there exists an $\alpha>0$ such 
that $E^\alpha$ satisfies~\eqref{basic}. Consequently, many more measures of entanglement are monogamous according to our definition.
We then provide a characterization for the disentangling condition in the form of an equality between the 
entanglement of formation (EoF) associated with the given entanglement measure (see Eq.~(\ref{eofmin}) below) 
and the entanglement of assistance (EoA)~\cite{DiVincenzo}, and discuss its relation to quantum Markov 
chains~\cite{HaydenJozaPetsWinter}. In addition, we characterize all states for which EoF equals to EoA 
when the measure of entanglement is taken to be the G-concurrence, and use that to show that the 
G-concurrence is monogamous. Finally,
we show that in the 2-dimensional case, the bipartite 2-qubit mixed states that can be expressed as 
the marginal of the 3-qubit $W$-state, are the only 2-qubit entangled states for which the 
EoF equals the EoA with respect to any measure of entanglement. 	
	

	Let $\mH^{A}\otimes\mH^B\equiv\mH^{AB}$ be a bipartite Hilbert space, and $\mS(\mH^{AB})\equiv\mS^{AB}$ 
	be the set of density matrices acting on $\mH^{AB}$. A function $E: \mS^{AB}\to\mbb{R}_{+}$ is called 
	a measure of entanglement if \textbf{(1)} $E(\sigma^{AB})=0$ for any separable density matrix $\sigma^{AB}\in\mS^{AB}$, 
	and \textbf{(2)} $E$ behaves monotonically under local operations and classical communications (LOCC). 
	That is, given an LOCC map $\Phi$ we have
	\be
	E\left(\Phi(\rho^{AB})\right)\leq E\left(\rho^{AB}\right)\;,\quad\;\;\forall\;\rho^{AB}\in\mS^{AB}\;.
	\ee
The measure is said to be \emph{faithful} if it is zero \emph{only} on separable states.
	
The map $\Phi$ is completely positive and trace preserving (CPTP). In general, LOCC can be stochastic, 
in the sense that $\rho^{AB}$ can be converted to $\sigma^{AB}_{j}$ with some probability $p_j$. 
In this case, the map from $\rho^{AB}$ to $\sigma_{j}^{AB}$ can not be described in general by a CPTP map. 
However, by introducing a `flag' system $A'$, we can view the ensemble $\{\sigma^{AB}_{j},p_j\}$ 
as a classical quantum state $\sigma^{A'AB}\equiv\sum_{j}p_j|j\lr j|^{A'}\otimes\sigma_{j}^{AB}$. 
Hence, if $\rho^{AB}$ can be converted by LOCC to $\sigma^{AB}_{j}$ with probability $p_j$, 
then there exists a CPTP LOCC map $\Phi$ such that $\Phi(\rho^{AB})=\sigma^{A'AB}$.
Therefore, the definition above of a measure of entanglement captures also probabilistic transformations. 
Particularly, $E$ must satisfy $E\left(\sigma^{A'AB}\right)\leq E\left(\rho^{AB}\right)$.

Almost all measures of entanglement studied in literature (although not all~\cite{Plenio2005}) satisfy
\be\label{emon}
E\left(\sigma^{A'AB}\right)=\sum_{j}p_jE(\sigma_{j}^{AB})\;,
\ee
which is very intuitive since $A'$ is just a classical system encoding the value of $j$. 
In this case the condition $E\left(\sigma^{A'AB}\right)\leq E\left(\rho^{AB}\right)$ 
becomes $\sum_{j}p_jE(\sigma_{j}^{AB})\leq E\left(\rho^{AB}\right)$. That is, LOCC can 
not increase entanglement on average. Measures of entanglement that satisfy this 
property are called entanglement monotones, and they can also be shown to be convex~\cite{Vidal2000}.
In the following definition we denote density matrices acting on a finite dimensional tripartite Hilbert space $\mH^{ABC}$
	by $\rho^{A|BC}$, where the vertical bar indicates the bipartite split across which we will measure the (bipartite) entanglement.
\begin{definition}\label{main}
Let $E$ be a measure of entanglement. $E$ is said to be \emph{monogamous} if for any $\rho^{A|BC}\in\mathcal{S}^{ABC}$ that satisfies
\be\label{cond}
	E(\rho^{A|BC})=E(\rho^{AB})	
\ee
we have that $E(\rho^{AC})=0$.
\end{definition}

The condition in~\eqref{cond} is a very strong one and typically is not satisfied by most tripartite 
states $\rho^{ABC}$. Following the terminology of~\cite{Hehuan} we call it here the ``disentangling condition". 
We will see below that states that saturate the strong subadditivity inequality for the von Neumann entropy 
(i.e. quantum Markov states) always satisfy this equality, for any entanglement monotone $E$.

Note that the condition in~\eqref{basic} is stronger than the one given in Definition~\ref{main}. Indeed,
if $E$ satisfies~\eqref{basic}, then any $\rho^{A|BC}$ that satisfies~\eqref{cond} must have
$E(\rho^{AC})=0$. At the same time, Definition~\ref{main} captures the essence of monogamy: 
that is, if system $A$ shares the maximum amount of entanglement with subsystem $B$, it is left with no entanglement to share with $C$.

In Definition~\ref{main} we do not invoke a particular monogamy relation such as~\eqref{basic}.
Instead, we propose a minimalist approach which is not quantitative, in which we only require what 
is essential from a measure of entanglement to be monogamous. Yet, this requirement is sufficient 
to generate a more quantitative monogamy relation:

\begin{theorem}
Let $E$ be a continuous measure of entanglement. Then, $E$ is monogamous 
according to Definition~\ref{main} if and only if there exists
$0<\alpha<\infty$ such that
	\be\label{power}
	E^\alpha(\rho^{A|BC})\geq E^\alpha(\rho^{AB})+E^\alpha(\rho^{AC})\;,	
	\ee
	for all $\rho^{ABC}\in\mathcal{S}^{ABC}$ with fixed $\dim\mH^{ABC}=d<\infty$.
\end{theorem}

We call the smallest possible value for $\alpha$ that satisfies Eq.~\eqref{power} in a given dimension
$d=\dim(\mH^{ABC})$,  the \emph{monogamy} exponent associated with a measure $E$, and denote it simply 
as $\alpha(E)$. In general, the monogamy exponent is hard to compute, and in the supplemental 
material we provide (along with the proofs of the theorems in this paper) a comprehensive list 
of known bounds for the monogamy exponent when $E$ is one of the measures of entanglement that were studied extensively in literature.

It is important to note that the relation given in~\eqref{power} is \emph{not} of the form
$
E(A|BC)\geq f\Big(E(A|B),E(A|C)\Big),
$
where $f$ is some function of two variables that satisfies certain conditions and is 
independent of $d$~\cite{Lancien}. This is because the monogamy exponent in~\eqref{power} 
depends on the dimension $d$, whereas $f$ as was used in~\cite{Lancien} is universal in the 
sense that it does not depend on the dimension. Therefore, if a measure of entanglement such 
as the entanglement of formation is not monogamous according to the class of relations given 
in~\cite{Lancien}, it does not necessarily mean that it is not monogamous according to our 
definition. Moreover, since our approach allows for dependence in the dimension, it avoids 
the issues raised in~\cite{Lancien} that measures of entanglement cannot be simultaneously 
monogamous and (geometrically) faithful (see~\cite{Lancien} for their definition of faithfulness).

In general, the class of all states $\rho^{ABC}$ that satisfy the disentangling
condition~\eqref{cond} depends on the choice of the entanglement measure $E$. However, 
there is a class of states that satisfy this condition for~\emph{any} choice of an entanglement 
monotone $E$. These are precisely the states that saturate the strong subadditivity of 
the von-Neumann entropy~\cite{HaydenJozaPetsWinter}.  For such states, the system $B$ 
Hilbert space $\mH^B$ must have a decomposition into a direct sum of tensor products
$
\mH^B=\bigoplus_{j}\mH^{B_{j}^{L}}\otimes\mH^{B_{j}^{R}}
$,
such that the state $\rho^{ABC}$ has the form
\be\label{Markov}
\rho^{ABC}=\bigoplus_{j}q_j\rho^{AB_{j}^{L}}\otimes\rho^{B_{j}^{R}C}\;,
\ee
where $q_j$ is a probability distribution~\cite{HaydenJozaPetsWinter}.

\begin{theorem}
Let $E$ be an entanglement monotone. Then, $E$ satisfies the disentangling condition~\eqref{cond} 
for all Markov quantum states $\rho^{ABC}$ of the form~\eqref{Markov}.
\end{theorem}

Note that a state of the form~\eqref{Markov} has a separable marginal state $\rho^{AC}$, 
and therefore the above theorem is consistent with Definition~\ref{main}.
Consequently, to check if an entanglement monotone is monogamous, one has to consider 
tripartite states that satisfy~\eqref{cond} but have a different form than~\eqref{Markov}. 
Perhaps such states do not exist for certain entanglement monotones. Indeed, partial 
results in this direction were proved recently in~\cite{Hehuan}. Particularly, it was 
shown that any pure tripartite entangled state $\psi^{ABC}$ with bipartite marginal 
state $\rho^{AB}$, that satisfies $N\left(\psi^{A|BC}\right)=N\left(\rho^{AB}\right)$, 
where $N$ is the negativity measure, must have the form $\psi^{ABC}=\psi^{AB^L}\otimes\psi^{B^RC}$. 
Note that this is precisely the form~\eqref{Markov} when $\rho^{ABC}$ is pure. 
Like the Negativity, we will see later that also the G-concurrence 
has this property for pure tripartite states.

Typically it is hard to check if a measure of entanglement is monogamous since 
the conditions in~\eqref{cond} involves mixed tripartite states. However, as 
we show below, it is sufficient to consider only \emph{pure} tripartite states 
in~\eqref{cond}, if the entanglement measure $E$ is defined on mixed states by a convex roof extension; that is,
\be
E_f\left(\rho^{AB}\right)\equiv\min\sum_{j=1}^{n}p_jE\left(|\psi_j\lr\psi_j|^{AB}\right)\;,\label{eofmin}
\ee
where the minimum is taken over all pure state decompositions of 
$\rho^{AB}=\sum_{j=1}^{n}p_j|\psi_j\lr\psi_j|^{AB}$.
We call $E_f$ the entanglement of formation (EoF) associated with $E$. 
In general, it could be that $E_f\neq E$ on mixed states like the convex roof extended Negativity is different than the Negativity itself.

\begin{theorem}\label{puremix}
Let $E_f$ be an entanglement monotone as above. If $E_f$ is monogamous 
(according to Definition~\ref{main}) on pure tripartite states in $\mH^{ABC}$, 
then it is also monogamous on tripartite mixed states acting on $\mH^{ABC}$.
\end{theorem}

From the above theorem and Corollary~\ref{eoaeof} below it follows that the convex 
roof extended negativity is monogamous, since it is monogamous for pure states~\cite{Hehuan}. 
Similarly, we will use it to show that the G-concurrence is monogamous. However, 
monogamy alone does not necessarily imply that a tripartite state is a Markov state if it satisfies the disentangling condition.
 In the following theorem, we provide yet another property of a tripartite state $\rho^{ABC}$ 
 that satisfies the disentangling condition. The property is that any LOCC protocol on such a 
 tripartite state can not increase on average the initial bipartite entanglement between $A$ and $B$. 
 The maximum such average of bipartite entanglement is known as \emph{entanglement of collaboration}~\cite{GourSpekkens}. It is defined by:
\be
E_{c}^{AB|C}\left(\rho^{ABC}\right)=\max\sum_{j=1}^{n}p_jE\left(\rho^{AB}_{j}\right)\;,\label{eocmixmax}
\ee
where $E$ is a given measure of bipartite entanglement, and the maximum is taken over all tripartite 
LOCC protocols yielding the bipartite state $\rho^{AB}_{j}$ with probability $p_j$. This measure of 
entanglement is closely related to EoA, denoted here by $E_{a}$, in which the optimization above is 
restricted to LOCC with one way classical communication from system $C$ to systems $A$ and $B$. 
Therefore, $E_c^{AB|C}\geq E_a$ and in~\cite{GourSpekkens} it was shown that this inequality can be strict.

\begin{theorem}\label{eoc}
Let $E$ be an entanglement monotone on bipartite mixed states, and let $\rho^{ABC}$ be a 
(possibly mixed) tripartite state satisfying the disentangling condition~(\ref{cond}). Then,
\be\label{strong}
E\left(\rho^{AB}\right)= E_{c}^{AB|C}\left(\rho^{ABC}\right)\;.
\ee
\end{theorem}

The condition in~\eqref{strong} is a very strong one. Particularly, it states that all measurements on system $C$ yield
the same average of bipartite entanglement between $A$ and $B$. Therefore, the entanglement 
of the marginal state $\rho^{AB}$ is \emph{resilient} to any quantum process or measurement on system $C$.
In the case that $\rho^{ABC}$ is pure, the EoA, $E_a$, depends only on the marginal $\rho^{AB}$, 
and we get the following corollary:
\begin{cor}\label{eoaeof}
Let $E$ be an entanglement monotone on bipartite states, and let $\rho^{ABC}$ be a pure tripartite 
state satisfying the disentangling condition~(\ref{cond}). Then,
\be
E\left(\rho^{AB}\right)=E_f(\rho^{AB})=E_a(\rho^{AB})\;,
\ee
where the EoF, $E_f$, is defined in~\eqref{eofmin}, and EoA, $E_a$, is also defined as in~\eqref{eofmin}
but with a maximum replacing the minimum.
\end{cor}

The equality of $E_f\left(\rho^{AB}\right)= E_a\left(\rho^{AB}\right)$ in corollary~\ref{eoaeof} 
implies that any pure state decomposition of $\rho^{AB}$ has the same average entanglement when 
measured by $E$. Unless $\rho^{AB}$ is pure, it is almost never satisfied. Nonetheless, 
there are non-trivial states (of measure zero) that do satisfy it. Such states have the following property:

\begin{theorem}\label{face}
Let $E$ be a measure of bipartite entanglement, and let $\rho^{AB}\in\mS^{AB}$ be a 
bipartite state with support subspace $\supp(\rho^{AB})$. If $E_f(\rho^{AB})=E_a(\rho^{AB})$ 
then for any $\sigma^{AB}\in\mS^{AB}$ with $\supp(\sigma^{AB})\subseteq\supp(\rho^{AB})$ we have $E_f(\sigma^{AB})=E_a(\sigma^{AB})$.
\end{theorem}
\begin{remark}
The theorem above follows from some of the results presented in~\cite{supUhlmann}, and for 
completeness we provide its full proof in the appendix.
\end{remark}
Theorem~\ref{face} demonstrates that the equality between the EoF and EoA corresponds to a 
property of the support of $\rho^{AB}$ rather than $\rho^{AB}$ itself. In the following theorem, 
we characterize the precise form of the support of $\rho^{AB}$ that yields such an equality. 
The entanglement monotone we are using is a generalization of the 
concurrence measure known as the G-concurrence~\cite{gour2}.

The G-concurrence is an entanglement monotone that on pure bipartite states is equal to the geometric-mean of the Schmidt coefficients.
On a (possibly unnormalized) vector $|x\ra\in\mbb{C}^d\otimes\mbb{C}^d$, it can be expressed as:
\be
G(|x\ra)=\left|\det(X)\right|^{2/d}\;;\quad\;\;|x\ra=X\otimes I|\phi_{+}\ra\;,
\ee
where $|\phi_{+}\ra=\sum_{i=1}^{d}|i\ra^A|i\ra^B$ is the maximally entangled state, 
and $X\in\mM_d(\mbb{C})$ is a $d\times d$ complex matrix. For mixed states the 
G-concurrence is defined in terms of the convex roof extension; that is, $G(\rho^{AB}):=G_{f}(\rho^{AB})$.

In the following theorem we denote by $\mN\subset\mM_d(\mbb{C})$, the \emph{nilpotent} cone, 
consisting of all $d\times d$ nilpotent complex matrices (i.e. $X\in\mN$ iff $X^k=0$ for some $1\leq k\leq d$).
While the set $\mN$ is not a vector space, it contains subspaces. For example, the set $\mT$ 
of all strictly upper triangular matrices (i.e. upper triangular matrices with zeros on the diagonal) 
is a $d(d-1)/2$-dimensional subspace  in $\mN$. From Gerstenhaber's theorem~\cite{Gerstenhaber1958} 
it follows that the largest dimension of a subspace in $\mN$ is $d(d-1)/2$, and if a 
subspace $\mN_0\subset\mN$ has this maximal dimension, then it must be similar to $\mT$ 
(i.e. their exists an invertible matrix $S$ such that $\mN_0=S\mT S^{-1}$).

\begin{theorem}\label{gcon}
Let $\rho^{AB}$ be a bipartite density matrix acting on $\mbb{C}^d\otimes\mbb{C}^d$ with rank $r$. Then,
\be\label{condition}
G(\rho^{AB})=G_a(\rho^{AB})>0\;,
\ee
 if and only if $r\leq 1+d(d-1)/2$, and there exists a full Schmidt rank state 
 $|x\ra^{AB}\in\mbb{C}^d\otimes\mbb{C}^d$, and a subspace $\mN_0\subset\mN$, with $\dim(\mN_0)=r-1$, such that
 \be\label{con2}
 \supp(\rho^{AB})= \{I\otimes Y|x\ra^{AB}\;\big|\;Y\in\mK\}\;,
 \ee
 where $\mK\equiv \{I\}\oplus\mN_0\equiv\{cI+N\;\big|\;c\in\mbb{C}\;;\;N\in\mN_0\}$.
 \end{theorem}
The direct sum above is consistent with the fact that a nilpotent matrix has a zero trace, 
so that it is orthogonal to the identity matrix in the Hilbert-Schmidt inner product.
The theorem above implies that the G-concurrence is monogamous  on pure tripartite states:

\begin{cor}
Let $|\psi^{ABC}\ra\in\mbb{C}^{d}\otimes\mbb{C}^{d}\otimes\mbb{C}^n$ be a pure tripartite state, 
with bipartite marginal $\rho^{AB}$. If $G(\psi^{A|BC})=G(\rho^{AB})>0$ 
then $|\psi\ra^{ABC}=|\chi\ra^{AB}|\varphi\ra^{C}$ for some bipartite state
$|\chi\ra^{AB}\in\mbb{C}^{d}\otimes\mbb{C}^{d}$ and a vector $|\varphi\ra^C\in\mbb{C}^n$.
\end{cor}
The above result is somewhat surprising since the G-concurrence is not a faithful measure of entanglement. Yet,
it states that the disentangling condition forces the marginal state $\rho^{AC}$ to be a product state. 
This is much stronger  than $G(\rho^{AC})=0$ (which can even hold for some entangled $\rho^{AC}$), 
and in particular, it states that $A$ and $C$ can not even share classical correlation. 
Combining the above corollary with Theorem~\ref{puremix} implies that the 
G-concurrence is monogamous on any mixed state that is acting on
$\mbb{C}^{d}\otimes\mbb{C}^{d}\otimes\mbb{C}^n$.
Finally, in the qubit case, Theorem~\ref{gcon} takes the following form:
 \begin{cor}
Let $\rho^{AB}$ be an entangled two qubit state with rank $r>1$, and let $E$ be any 
injective (up to local unitaries) measure of pure two qubit entanglement.
Then,
\begin{enumerate}
\item  If $E_f(\rho^{AB})=E_a(\rho^{AB})$ then $r=2$, and $\rho^{AB}=\tr_C|W\lr W|^{ABC}$ 
is the 2-qubit marginal of $|W\ra=\lambda_1|100\ra+\lambda_2|010\ra+\lambda_3|001\ra+\lambda_4|000\ra$, $\lambda_i\in\mbb{C}$.
\item  Conversely, if $\rho^{AB}$ is a marginal of a state in the W-class then
$C_f(\rho^{AB})=C_a(\rho^{AB})$, where $C$ is the concurrence.
\end{enumerate}
\end{cor}

\begin{remark}
The second part of the theorem implies that also $E_f(\rho^{AB})=E_a(\rho^{AB})$, 
for any $\rho^{AB}$ that is a marginal of a W-state, and any measure $E$, 
that can be expressed as a convex function of the concurrence $C$~\cite{Woo98}.
\end{remark}

In conclusions, we introduced a new definition for a monogamous measure of entanglement. 
Our definition involves an equality (the disentangling condition~\eqref{cond}) 
rather than the inequality~\eqref{basic}. Yet, we showed that our notion of monogamy can 
reproduce monogamy relations like in~\eqref{basic} with a small change that the measure 
$E$ is replaced by $E^\alpha$ for some exponent $\alpha>0$. We then showed that convex 
roof based entanglement monotones of the form~\eqref{eofmin} are monogamous iff they are monogamous on pure tripartite states,
and showed further that the disentangling condition in~\eqref{cond} holds for any 
entanglement monotone if $\rho^{ABC}$ is a quantum Markov state. While it is left open 
if the converse is also true (at least for some entanglement monotones), we were able to 
show that for the G-concurrence, the only \emph{pure} tripartite states that satisfy 
the disentangling condition~\eqref{cond} are Markov states.
In addition, we related the disentangling condition to states that have the same average 
entanglement for all convex pure state decompositions, and found a characterization of 
such states in Theorem~\ref{face} (for general measures of entanglement), and a complete 
characterization in Theorem~\ref{gcon} (for the G-concurrence).  
Clearly, much more is left to investigate along these lines.

	\begin{acknowledgements}
		This work was completed while Y.G was visiting
		the Institute of Quantum Science and
		Technology of the University of Calgary
		under the
		support of the China Scholarship Council under Grant No. 201608140008.
		Y.G thanks Professor
		C. Simon
		and Professor G. Gour for their hospitality,
		and thanks Sumit Goswami for helpful discussions.
		G.G research is supported by the Natural Sciences and Engineering Research Council of Canada (NSERC).
		Y.G is supported by the National Natural
		Science Foundation of China under Grant No. 11301312 and
		the Natural Science Foundation of Shanxi
		under Grant No. 201701D121001.
		
	\end{acknowledgements}
	

\begin{titlepage}
\center{\large\textbf{Supplementary Material\\Monogamy of entanglement without inequalities}\\}
\center{~{ }\\}

\end{titlepage}

\onecolumngrid

\appendix

	\section{The Monogamy exponent}
	
\noindent\textbf{Theorem 1. }{\it
Let $E$ be a continuous measure of entanglement. Then, $E$ is monogamous 
according to Definition~\ref{main} if and only if there exists
$0<\alpha<\infty$ such that
	\be\label{power2}
	E^\alpha(\rho^{A|BC})\geq E^\alpha(\rho^{AB})+E^\alpha(\rho^{AC})\;,	
	\ee
	for all $\rho^{ABC}\in\mathcal{S}^{ABC}$ with fixed $\dim\mH^{ABC}=d<\infty$.
}

\begin{proof}
Let $E$ be a monogamous measure of entanglement according to Def.\ref{main}.
	Since $E$ is a measure of entanglement,
	it is non-increasing under partial traces, and therefore
	$E(\rho^{A|BC})\geq \max\{E(\rho^{AB}),E(\rho^{AC})\}$ for
	any state $\rho^{A|BC}\in\mathcal{S}^{ABC}$.
	We assume $E(\rho^{A|BC})>0$ and set $x_1\equiv E(\rho^{AB})/E(\rho^{A|BC})$
	and $x_2\equiv E(\rho^{AC})/E(\rho^{A|BC})$.	Clearly,
	there exists $\gamma>0$ such that
	\be\label{equality}
	x_1^\gamma+x_2^\gamma\leq 1\;,
	\ee
	since either $x_j^\gamma\rightarrow 0$ when $\gamma$ increases, or if 
	$x_1=1$ then by assumption $x_2=0$ and vise versa. We denote by
	$f(\rho^{ABC})$ the smallest value of $\gamma$ that achieves equality in~\eqref{equality}.
	Since $E$ is continuous, so is $f$, and the compactness of $\mathcal{S}^{ABC}$ gives:
	\be\label{optimal}
	\alpha\equiv\max_{\rho^{ABC}\in\mathcal{S}^{ABC}}f(\rho^{ABC})<\infty\;.
	\ee
	By definition, $\alpha$ satisfies the condition in~(\ref{power}).
\end{proof}

	As discussed in the paper, the expression for $\alpha$ in~\eqref{optimal} is optimal 
	in the sense that it provides the smallest possible value for $\alpha$ that 
	satisfies Eq.~\eqref{power}. This monogamy exponent is a function of the measure 
	$E$, and we denote it by $\alpha(E)$.  It may depend also on the dimension 
	$d\equiv\dim(\mH^{ABC})$ (see in Table~\ref{tab:table1}), and, in general, 
	is hard to compute especially in higher dimensions~\cite{supHehuan,supLancien,
		supAllen,supEltschka,supAudenaert,supChengshuming}. By definition, $\alpha(E)$ 
	can only increase with $d$ (e.g.
	Table~\ref{tab:table1}).
Table~\ref{tab:table1} indicates that almost any entanglement measure is 
monogamous at least for multi-qubit systems. In addition, almost all the
	entanglement measures studied in the literature are continuous, and in particular  $C$, $N$, $N_{cr}$,
	$E_f$, $\tau$,  $T_q$, $R_\alpha$ and $E_r$ are all continuous~\cite{supGuo2013qip,supGuo2013csb,
		supDonald1999pla,supGuo}.

		\begin{table}
		\caption{\label{tab:table1}A comparison of the monogamy exponent of
			several entanglement measures.
			We denote the one-way distillable entanglement, concurrence, negativity,
			convex roof extended negativity, entanglement of formation 
			(the original one defined in~\cite{supHillWotters}),
			tangle, squashed entanglement, Tsallis-$q$ entanglement and
			R\'{e}nyi-$\alpha$ entanglement by $E_d$,
			$C$, $N$,$N_{cr}$, $E_f$, $\tau$, $E_{sq}$, $T_q$ and $R_\alpha$, respectively.}
		\begin{ruledtabular}
			\begin{center}
			\begin{tabular}{cccc}\hline
				$E$& $\alpha(E)$ & System & Reference \\ \hline
				$E_d$ & $\leq1$& any system & \cite{supKoashi}\\
				$C$ & $2$& $2^{\otimes 3}$ & \cite{supCoffman}\footnotemark[1]\\
				& $\leq\sqrt{2}$& $2^{\otimes n}$ & \cite{supZhuxuena2014pra}\\
				&   $\leq 2$    & $2\otimes 2\otimes2^m$ &\cite{supOsborne}\\
				&   $\leq 2$    & $2^{\otimes n}$ &\cite{supOsborne}\\
				&   $>2$        & $3^{\otimes 3}$ & \cite{supOuyongcheng}\\
				&   $\leq 2$    & $2\otimes 2\otimes4$ &\cite{supRenxijun} \\
				&   $>3$    & $3\otimes 2\otimes2$ &\cite{supKim} \\
				$N$& $\leq 2$ & $2^{\otimes n}$ &\cite{supOuyongcheng2007pra2,supLuo}\footnotemark[2]\\
				& $\leq 2$ & $2\otimes 2\otimes2^m$ &\cite{supLuo}\footnotemark[2]\\
				& $\leq 2$ & $d\otimes d\otimes d$,$d=2,3,4$ &\cite{supHehuan}\\
				$N_{cr}$ & $\leq 2$ & $2^{\otimes n}$ & \cite{supKim2009,supLuo,supChoi}\\
				$E_f$& $\leq 2$ & $2^{\otimes n}$ & \cite{supKumar,supBai}\\
				& $\leq \sqrt{2}$ & $2^{\otimes n}$ & \cite{supZhuxuena2014pra}\\
				& $>1$ & $2\otimes 2\otimes 2$ & \cite{supBai,supOliveira2014pra}\\
				$\tau$ &   $\leq 1$    & $2\otimes 2\otimes4$ &\cite{supRenxijun}\\
				$E_{sq}$& $\leq 1$& any system &\cite{supKoashi}\\
				$T_{q}$, $2\leq q\leq 3$& $\leq 1$& $2^{\otimes n}$ &\cite{supKim2016pra}\\
				$T_{q}$& $\leq 2$& $2\otimes 2\otimes2^m$ &\cite{supLuo2016pra}\footnotemark[3]\\
				$R_{\alpha}$, $\alpha\geq2$& $\leq 1$& $2^{\otimes n}$ &\cite{supKim2010jpa,supCornelio}\\
				$R_{\alpha}$, $\alpha\geq\frac{\sqrt{7}-1}{2}$& $\leq 2$& $2^{\otimes n}$ &\cite{supSong}\footnotemark[4]\\
				\hline
			\end{tabular}
		     \end{center}
		\end{ruledtabular}
	    \footnotetext[1]{$\alpha(C)\leq2$ was shown in~\cite{supCoffman},
	    and the equality $\alpha(C)=2$ follows from the saturation by $W$ states (see, for example, Corollary 9).}
		\footnotetext[2]{For pure states.}
		\footnotetext[3]{For mixed states, and $q\in[\frac{5-\sqrt{13}}{2},2]\cup[3,\frac{5+\sqrt{13}}{2}]$.}
		\footnotetext[4]{For mixed states.}
	\end{table}

\section{Quantum Markov States and Monogamy of Entanglement}

Recall that quantum Markov states are states that saturate the strong subadditivity 
of the von-Neumann entropy. That is, they saturate the inequality:
\be
S(\rho^{AB})+S(\rho^{BC})\geq S(\rho^{ABC})+S(\rho^B)\;,
\ee
where $S(\rho)=-\tr\left[\rho\log\rho\right]$ is the von-Neumann entropy. 
In~\cite{supHaydenJozaPetsWinter} it was shown that the inequality above 
is saturated if and only if the Hilbert space of system $B$, $\mH^B$, can be decomposed into a direct sum of tensor products
\be
\mH^B=\bigoplus_{j}\mH^{B_{j}^{L}}\otimes\mH^{B_{j}^{R}}
\ee
such that the state $\rho^{ABC}$ has the form
\be\label{aMarkov}
\rho^{ABC}=\bigoplus_{j}q_j\rho^{AB_{j}^{L}}\otimes\rho^{B_{j}^{R}C}\;,
\ee
where $q_j$ is a probability distribution.\\

\noindent\textbf{Theorem 2. }{\it
Let $E$ be an entanglement monotone. Then, $E$ satisfies the disentangling 
condition~\eqref{cond} for all entangled Markov quantum states $\rho^{ABC}$ of the form~\eqref{aMarkov}.}

\begin{proof}
	
	Since local ancillary systems are free in entanglement theory, one can append 
	an ancillary system $B'$ that encodes the orthogonality of the 
	subspaces $\mH^{B_{j}^{L}}\otimes\mH^{B_{j}^{R}}$. This can be done with an 
	isometry that maps states in $\mH^{B_{j}^{L}}\otimes\mH^{B_{j}^{R}}$ to states 
	in $\mH^{B^{L}}\otimes\mH^{B^{R}}\otimes|j\lr j|^{B'}$, where systems $B^L$ and 
	$B^R$ have dimensions $\max_j\dim\left(\mH^{B_{j}^{L}}\right)$ and  $\max_j\dim\left(\mH^{B_{j}^{R}}\right)$, respectively.
Therefore, w.l.o.g. we can write the above state as
\be\label{mainform}
\rho^{ABC}=\sum_{j}q_j\;\rho^{AB^{L}}_{j}\otimes|j\lr j|^{B'}\otimes\rho^{B^{R}C}_{j}\;.
\ee
Now, note that with any entanglement monotone $E$, the entanglement between $A$ and $BC$ is measured by:
\be
E\left(\rho^{A|BC}\right)=\sum_{j}q_j E\left(\rho^{A|B_{j}^{L}}\otimes\rho^{B_{j}^{R}C}\right)=
\sum_{j}q_j E\left(\rho^{AB_{j}^{L}}\right)\;,
\ee
where in the first equality we used the property~\eqref{emon} of entanglement monotones.
Similarly, the entanglement between $A$ and $B$ is measured by
\be
E\left(\rho^{AB}\right)=\sum_{j}q_j E\left(\rho^{AB_{j}^{L}}\otimes\rho^{B_{j}^{R}}\right)=
\sum_{j}q_j E\left(\rho^{AB_{j}^{L}}\right)\;.
\ee
We therefore obtain~\eqref{cond} as long as $E\left(\rho^{A|B_{j}^{L}}\right)>0$ for some $j$ for which $q_j>0$.
This completes the proof.
\end{proof}

\section{Monogamy of entanglement: pure vs mixed tripartite states }

As discussed in the paper, it is typically hard to check if a measure of entanglement is 
monogamous since the condition in~\eqref{cond} involves mixed tripartite states. 
On the other hand, it is significantly simpler to check the disentangling condition 
if $\rho^{ABC}$ that appears in the disentangling condition~\eqref{cond} is pure. 
We say that $E$ is monogamous on pure states if for any pure tripartite state $\rho^{ABC}$ that satisfies~\eqref{cond}, $E(\rho^{AC})=0$.
In the theorem below we shown that sometimes if $E$ is monogamous on pure states 
it is also monogamous on mixed states (that is, it is monogamous according to Def.~\ref{main}).

For any entanglement monotone $E$ on the set of bipartite density matrices, $\mS^{AB}$, 
we define a corresponding entanglement of formation measure, $E_f$ which is defined by the following convex roof extension:
\be\label{aeofmin}
E_f\left(\rho^{AB}\right)\equiv\min\sum_{j=1}^{n}p_jE\left(|\psi_j\lr\psi_j|^{AB}\right)\;,
\ee
where the minimum is taken over all pure state decompositions of $\rho^{AB}=\sum_{j=1}^{n}p_j|\psi_j\lr\psi_j|^{AB}$.
Clearly, $E=E_f$ on pure bipartite states, but on mixed states they can be different, 
like the convex roof extended Negativity is different from the Negativity itself. 
Since we assume that $E$ is entanglement monotone it is convex so
$E(\rho^{AB})\leq E_{f}(\rho^{AB})$ for all $\rho^{AB}\in\mS^{AB}$. The corresponding 
entanglement of formation of a given entanglement monotone, is itself an entanglement monotone, and has the following remarkable property.\\

\noindent\textbf{Theorem 3. } {\it
Let $E$ be an entanglement monotone, and let $E_f$ be its corresponding entanglement of 
formation~\eqref{aeofmin}. If $E_f$ is monogamous (according to Definition~\ref{main}) 
on pure tripartite states in $\mH^{ABC}$, then it is also monogamous on tripartite mixed states acting on $\mH^{ABC}$.
}

\begin{proof}
Let $\rho^{A|BC}=\sum_{j}p_j|\psi_j\lr\psi_j|^{ABC}$ be a tripartite state acting on 
$\mH^{ABC}$ with $\{p_j,|\psi_j\ra^{ABC}\}$ being the optimal decomposition such that
\be
E_f(\rho^{A|BC})=\sum_{j}p_jE_f\left(|\psi_j\ra^{A|BC}\right)\;.
\ee
We also assume w.l.o.g. that $p_j>0$.
Now, suppose $E_f(\rho^{A|BC})=E_f(\rho^{AB})$, and denote $\rho_{j}^{AB}\equiv\tr_{C}|\psi_j\lr\psi_j|^{ABC}$.
Since discarding a subsystem can only decrease the entanglement, we get
\be
\sum_{j}p_jE_f\left(|\psi_j\ra^{A|BC}\right)\geq \sum_{j}p_jE_f\left(\rho_{j}^{AB}\right)\geq E_f(\rho^{AB})\;,
\ee
where the last inequality follows from the convexity of $E_f$ and the fact that $\rho^{AB}=\sum_jp_j \rho_{j}^{AB}$.
However, all the inequalities above must be equalities since $E_f(\rho^{A|BC})=E_f(\rho^{AB})$. In particular, we get
\be
\sum_{j}p_jE_f\left(|\psi_j\ra^{A|BC}\right)= \sum_{j}p_jE_f\left(\rho_{j}^{AB}\right)\;.
\ee
This in turn implies that $E_f\left(|\psi_j\ra^{A|BC}\right)=E_f\left(\rho_{j}^{AB}\right)$ for each $j$ since
$E_f\left(|\psi_j\ra^{A|BC}\right)\geq E_f\left(\rho_{j}^{AB}\right)$ for each $j$ (i.e. 
tracing out subsystem cannot increase entanglement). Since we assume that $E_f$ is 
monogamous on pure tripartite states, we conclude that for each $j$,
$E_f(\rho_{j}^{AC})=0$, where $\rho_{j}^{AC}\equiv\tr_B |\psi_j\lr\psi_j|^{ABC}$. 
Hence, $E_f(\rho^{AC})=0$ since $\rho^{AC}=\sum_jp_j \rho_{j}^{AC}$ and $E_f$ is convex.
\end{proof}

\section{Entanglement of Collaboration and Monogamy of Entanglement}

Monogamy of entanglement is closely related to entanglement of collaboration. 
Given a measure of bipartite entanglement $E$, its corresponding entanglement of 
collaboration, $E_c^{AB|C}$, is a measure of entanglement on tripartite mixed states, $\rho^{ABC}$, given by~\cite{supGourSpekkens}:
\be
E_c^{AB|C}\left(\rho^{ABC}\right)=\max\sum_{j=1}^{n}p_jE\left(\rho^{AB}_{j}\right)\;,\label{aeocmixmax}
\ee
where the maximum is taken over all tripartite LOCC protocols yielding the bipartite 
state $\rho^{AB}_{j}$ with probability $p_j$.
The following theorem demonstrates the connection between the disentangling 
condition and entanglement of collaboration.\\

\noindent\textbf{Theorem 4. }{\it
Let $E$ be an entanglement monotone on bipartite mixed states, and let $\rho^{ABC}$ be 
a (possibly mixed) tripartite state satisfying the disentangling condition~(\ref{cond}). Then,
\be\label{astrong}
E\left(\rho^{AB}\right)= E_c^{AB|C}\left(\rho^{ABC}\right)\;.
\ee
}

\begin{proof}
Let $\{\rho^{AB}_{j},\;p_j\}$ be the optimal ensemble in~\eqref{eocmixmax} obtained by 
LOCC on $\rho^{ABC}$. Since $E$ is a bipartite entanglement monotone, it does not increase on average:
\be
E\left(\rho^{A|BC}\right)\geq \sum_{j}p_jE(\rho^{AB}_{j})=E_c^{AB|C}\left(\rho^{ABC}\right)\;.
\ee
On the other hand, by definition $E(\rho^{AB})\leq E_c^{AB|C}\left(\rho^{ABC}\right)$, 
so that together with~\eqref{cond} we get~\eqref{strong}.
\end{proof}

Entanglement of collaboration is different from entanglement of assistance, $E_a$, 
in which the optimization in~\eqref{aeocmixmax} is restricted to LOCC of the following form: 
Charlie (system C) performs a measurement, and communicates the outcome $j$ to Alice and Bob. 
In~\cite{supGourSpekkens} the following LOCC protocol was considered: Alice performs a measurement, 
then sending the outcome to Charlie, and then Charlie performs his measurement, and sends back 
the result to Alice and Bob. It was shown that in such a scenario it is possible to increase 
the average entanglement between systems $A$ and $B$ to a value beyond the average 
entanglement that can be achieved if only Charlie performed a measurement. Therefore, 
$E_c^{AB|C}$ can be strictly larger than $E_a$, and in general, $E_c^{AB|C}\geq E_a$. 
However, if $\rho^{ABC}$ satisfies the disentangling condition then we must have 
$E_c^{AB|C}=E_a$. Indeed, if $\rho^{ABC}$ satisfies~\eqref{cond} we get
\be
E_a(\rho^{ABC})\geq E(\rho^{AB})=E\left(\rho^{A|BC}\right)=E_c^{AB|C}\left(\rho^{ABC}\right)\;.
\ee
Therefore, one can replace $E_c^{AB|C}$ in~\eqref{astrong} with $E_a$, which may be 
convenient since $E_a$ is somewhat a simpler measure than $E_c^{AB|C}$. Note however 
that we left $E_c^{AB|C}$ in~\eqref{astrong} since 
$E\left(\rho^{A|BC}\right)=E_c^{AB|C}\left(\rho^{ABC}\right)$ 
implies $E\left(\rho^{A|BC}\right)=E_{a}\left(\rho^{ABC}\right)$ but not vice versa.

\subsection{When Entanglement of Formation equals Entanglement of Assistance?}

An immediate consequence of Theorem~\ref{eoc} above is that if $\rho^{ABC}$ 
is a pure state that satisfies the disentangling condition then the entanglement 
of formation of $\rho^{AB}$ must be equal to its entanglement of assistance.\\

\noindent\textbf{Corollary 5. }{\it
Let $E$ be an entanglement monotone on bipartite states, and let $\rho^{ABC}$ 
be a pure tripartite state satisfying the disentangling condition~(\ref{cond}). Then,
\be\label{aeq}
E\left(\rho^{AB}\right)=E_f(\rho^{AB})=E_a(\rho^{AB})\;,
\ee
where the entanglement of formation, $E_f$, is defined in~\eqref{aeofmin}, 
and the entanglement of assistance, $E_a$, is also defined as in~\eqref{aeofmin}
but with a maximum replacing the minimum.
}

\begin{proof}
The proof follows straightforwardly from Theorem~\eqref{eoc} recalling that
\be
E\left(\rho^{AB}\right)\leq E_f(\rho^{AB})\leq E_a(\rho^{AB})\leq E_c^{AB|C}\left(\rho^{ABC}\right)=E\left(\rho^{A|BC}\right)\;,
\ee
where the first inequality follows from the fact that $E$ is an entanglement monotone, 
and the last equality from Theorem~\eqref{eoc}. Therefore, all the inequalities 
above are equalities since we assume the disentangling condition
$E\left(\rho^{A|BC}\right)=E\left(\rho^{AB}\right)$. This completes the proof.
\end{proof}

In the following theorem we show that the equality between the entanglement of 
formation and entanglement of assistance is a property of the support space of the the bipartite state in question.\\

\noindent\textbf{Theorem 6. }{\it
Let $E$ be a measure of bipartite entanglement, and let $\rho^{AB}\in\mS^{AB}$ 
be a bipartite state with support subspace $\supp(\rho^{AB})$. If 
$E_f(\rho^{AB})=E_a(\rho^{AB})$ then for any $\sigma^{AB}\in\mS^{AB}$ 
with $\supp(\sigma^{AB})\subseteq\supp(\rho^{AB})$ we have $E_f(\sigma^{AB})=E_a(\sigma^{AB})$.
}

\begin{proof}
In the proof we use some of the ideas introduced in~\cite{supUhlmann}. Suppose
$\rho^{AB}=\sum_{j}p_j\rho_{j}^{AB}$, where $\{p_j\}$ are (non-zero) 
probabilities and $\rho^{AB}_{j}\in\mS^{AB}$. Then, the condition 
$E_f(\rho^{AB})=E_a(\rho^{AB})$ together with the convexity (concavity) of $E_f$ ($E_a$) gives
\be
\sum_{j}p_jE_f(\rho^{AB}_{j})\geq E_f(\rho^{AB})=E_a(\rho^{AB})\geq \sum_{j}p_jE_{a}(\rho^{AB}_{j})\;.
\ee
But since for all $j$ we also have $E_f(\rho^{AB}_{j})\leq E_a(\rho^{AB}_{j})$, 
we get that  $E_f(\rho^{AB}_{j})= E_a(\rho^{AB}_{j})$ for all $j$. Let $\mF(\rho^{AB})$ 
be the set of \emph{all} density matrices in $\mS^{AB}$ that appear in a convex 
decomposition of $\rho^{AB}$. In convex analysis, $\mF(\rho^{AB})$ is called a 
\emph{face} of $\mS^{AB}$. Now, from the argument above we have that if 
$\sigma^{AB}\in\mF(\rho^{AB})$ then $E_f(\sigma^{AB})=E_a(\sigma^{AB})$. 
On the other hand, for any $\sigma^{AB}\in\mS^{AB}$ with the property that 
$\supp(\sigma^{AB})\subset\supp(\rho^{AB})$ there exists a small enough 
$\epsilon>0$ such that $\rho^{AB}-\epsilon\sigma^{AB}\geq 0$. Denoting by
$\gamma^{AB}\equiv \left(\rho^{AB}-\epsilon\sigma^{AB}\right)/(1-\epsilon)$ 
we get that $\gamma^{AB}\in\mS^{AB}$ and $\rho^{AB}$ can be expressed as 
the convex combination $\rho^{AB}=\epsilon\sigma^{AB}+(1-\epsilon)\gamma^{AB}$. 
We therefore must have $E_f(\sigma^{AB})=E_a(\sigma^{AB})$. This completes the proof.
\end{proof}

In the proof above we called $\mF(\rho^{AB})$ a face.
A face $\mF$ of $\mS^{AB}$ is a \emph{convex} subset of $\mS^{AB}$ that satisfies the following property:
if $t\rho_1+(1-t)\rho_2\in\mF$ for some $\rho_1,\rho_2\in\mS^{AB}$ and $0<t<1$, then $\rho_1,\rho_2\in\mF$.
To see that $\mF(\rho^{AB})$ is a face of $\mS^{AB}$, we first show that it is 
convex. Indeed, let $\sigma\equiv t\sigma_1+(1-t)\sigma_2$ for some $t\in[0,1]$ 
and $\sigma_1,\sigma_2\in\mF(\rho^{AB})$. Since $\sigma_1,\sigma_2\in\mF(\rho^{AB})$ 
there exists $p,q\in(0,1]$ and $\gamma_1,\gamma_2\in\mS^{AB}$ such that
\be
\rho^{AB}=p\sigma_1+(1-p)\gamma_1=q\sigma_2+(1-q)\gamma_2\;.
\ee
The first equality implies that $\frac{t}{p}\rho^{AB}=t\sigma_1+\frac{t(1-p)}{p}\gamma_1$, 
and the second equality gives
$\frac{1-t}{q}\rho^{AB}=(1-t)\sigma_2+\frac{(1-t)(1-q)}{q}\gamma_2$. Therefore,
\be
\left(\frac{t}{p}+\frac{1-t}{q}\right)\rho^{AB}=\sigma^{AB}+\frac{t(1-p)}{p}\gamma_1+\frac{(1-t)(1-q)}{q}\gamma_2\;.
\ee
After dividing by $\frac{t}{p}+\frac{1-t}{q}$, we can see that $\sigma^{AB}$ appears 
in a convex combination of $\rho^{AB}$. Therefore, $\mF(\rho^{AB})$ is convex. 
To complete the proof that $\mF(\rho^{AB})$ is a face, note that if 
$\tau\equiv t\rho_1+(1-t)\rho_2\in\mF$ for some $\rho_1,\rho_2\in\mS^{AB}$ and $0<t<1$ 
then clearly $\rho_1,\rho_2\in\mF(\rho^{AB})$.

Note that the condition $\supp(\sigma^{AB})\subseteq\supp(\rho^{AB})$ is equivalent to $\sigma^{AB}\in\mF(\rho^{AB})$.
The precise form of the support space of a bipartite state $\rho^{AB}$ 
that satisfies $E_f(\rho^{AB})=E_a(\rho^{AB})$ depends on the measure of entanglement $E$. 
In the following sections we find it precisely for the case where $E$ is the G-concurrence, 
and we use it to show that the G-concurrence is monogamous.

\subsection{Monogamy of the G-concurrence}

Any pure bipartite state, $|x\ra\in\mbb{C}^d\otimes\mbb{C}^d$, can be written as:
\be\label{iso2}
|x\ra=X\otimes I|\phi_{+}\ra\quad\text{where}\quad|\phi_{+}\ra=\sum_{i=1}^{d}|i\ra^A|i\ra^B\;,
\ee
and $X$ is a $d\times d$ complex matrix. The relation above between a complex matrix 
$X\in\mM_d(\mbb{C})$ and a bipartite vector $|x\ra\in\mbb{C}^d\otimes\mbb{C}^d$ defines 
an isomorphism between $\mM_d(\mbb{C})$ and $\mbb{C}^d\otimes\mbb{C}^d$.
Using this isomorphism, in the remaining of this section we will view interchangeably  
the support of a density matrix both as a subspace of $\mM_d(\mbb{C})$ or as a 
subspace of $\mbb{C}^d\otimes\mbb{C}^d$, depending on the context. We will use 
capital letters $X,Y,Z,W$ for matrices in $\mM_d(\mbb{C})$, and use lower case 
letters $|x\ra,|y\ra,|z\ra,|w\ra$ to denote their corresponding bipartite vectors.

With these notations, the G-concurrence of $|x\ra$, which is the geometric 
mean of the Schmidt coefficients of $|x\ra$,
can be expressed as:
\be\label{defgcon}
G(|x\ra)=\left|\det(X)\right|^{2/d}\;,
\ee
and for mixed states it is defined in terms of the convex roof extension; that is, 
$G(\rho^{AB}):=G_{f}(\rho^{AB})$ for any $\rho^{AB}\in\mS^{AB}$. Note that the 
G-concurrence is homogeneous, and in particular, $G(c|x\ra)=|c|^2G(|x\ra)$.   
We start in proving the following Lemma:

\noindent\textbf{Lemma: }{\it
Let $\rho^{AB}$ be a bipartite density matrix acting on $\mbb{C}^d\otimes\mbb{C}^d$ 
with $G(\rho^{AB})>0$. Then, there exists a pure state decomposition of $\rho^{AB}$ with the following properties:
\be\label{w2}
\rho^{AB}=\sum_{j=1}^{r}|w_j\lr w_j|\;,\quad\;G(|w_j\ra)=0\;,\;\;\forall\;j=2,...,r,
\ee
where $r$ is the rank of $\rho^{AB}$ and $|w_j\ra$ are sub-normalized vectors in $\mbb{C}^d\otimes\mbb{C}^d$
(i.e. vectors with norm no greater than 1).
}

\begin{proof}
Let $\rho^{AB}=\sum_{j=1}^{r}|x_j\lr x_j|$ be the spectral decomposition of $\rho^{AB}$ 
with $|x_j\ra$ being the sub-normalized eigenvectors of $\rho^{AB}$. 
Clearly, $G(|x_j\ra)>0$ for at least one $j$. Therefore, w.l.o.g. we assume
$G(|x_1\ra)>0$.   Let $\mK_2\subset\supp(\rho^{AB})$ be the two dimensional 
subspace spanned by $X_1$ and $X_2$.
We first show that $\mK_2$ contains a matrix with zero determinant. 
Indeed, if $\det(X_2)=0$ we are done. Otherwise,
for any $\lambda\in\mbb{C}$ we get
\begin{align}
G(|x_1\ra+\lambda|x_2\ra)&=\left|\det(X_1+\lambda X_2)\right|^{2/d}\nonumber\\
&=\left|\det(X_2)\right|^{2/d}\left|\det(X_1X_{2}^{-1}+\lambda I)\right|^{2/d}\;.
\end{align}
Note that $\det(X_1X_{2}^{-1}+\lambda I)$ is a polynomial of degree $d$ in 
$\lambda$ and must have at least one complex root. Therefore, 
there exists $\lambda=\lambda_0\neq 0$ such that $\det(X_1+\lambda_0 X_2)=0$.
This completes the assertion that $\mK_2$ contains a matrix with zero determinant.

Next, denote
$a\equiv\frac{1}{\sqrt{1+|\lambda_0|^2}}$ and $b\equiv \frac{\lambda_0}{\sqrt{1+|\lambda_0|^2}}$, so that
$|a|^2+|b|^2=1$, and the vectors $|w_2\ra\equiv a|x_1\ra+b|x_2\ra$ and $|y\ra\equiv \bar{b}|x_1\ra-\bar{a}|x_2\ra$
satisfy
\be
|x_1\lr x_1|+|x_2\lr x_2|=|y\lr y|+|w_2\lr w_2|
\ee
with $G(|w_2\ra)=0$ (by construction, $W_2$ is proportional to 
$X_1+\lambda_0 X_2$ and therefore has zero determinant). We can therefore write
\be
\rho^{AB}=|w_2\lr w_2|+|y\lr y|+\sum_{j=3}^{r}|x_j\lr x_j|\;.
\ee
Now, if $G(|y\ra)=0$ then we denote it as $|w_3\ra$ and pick from $\{|x_j\ra\}_{j=3}^{r}$ 
another state that does not have a vanishing G-concurrence. We therefore assume $G(|y\ra)\neq 0$ 
and from the same arguments as above we conclude that 
$|y\lr y|+|x_3\lr x_3|=|\tilde{y}\lr \tilde{y}|+|w_3\lr w_3|$ for some vectors 
$|\tilde{y}\ra$ and $|w_3\ra$, with $G(|w_3\ra)=0$. By replacing $|y\ra$ and 
$|x_3\ra$ with $|\tilde{y}\ra$ and $|w_3\ra$, and repeating the process 
we arrive at the desired decomposition of $\rho^{AB}$.
\end{proof}

In the following theorem we denote by $\mN\subset\mM_d(\mbb{C})$, the 
\emph{nilpotent} cone, consisting of all $d\times d$ nilpotent 
complex matrices (i.e. $X\in\mN$ iff $X^k=0$ for some $1\leq k\leq d$).
While the set $\mN$ is not a vector space, it contains subspaces. 
For example, the set $\mT$ of all strictly upper triangular matrices 
(i.e. upper triangular matrices with zeros on the diagonal) is a $d(d-1)/2$-dimensional 
subspace  in $\mN$. From Gerstenhaber's theorem~\cite{supGerstenhaber1958} it 
follows that the largest dimension of a subspace in $\mN$ is $d(d-1)/2$, and 
if a subspace $\mN_0\subset\mN$ has this maximal dimension, then it must be 
similar to $\mT$ (i.e. their exists an invertible matrix $S$ such that $\mN_0=S\mT S^{-1}$).\\

\noindent\textbf{Theorem 7. }{\it
Let $\rho^{AB}$ be a bipartite density matrix acting on $\mbb{C}^d\otimes\mbb{C}^d$, 
and suppose it has rank $r>1$. Then,
\be\label{acondition}
G(\rho^{AB})=G_a(\rho^{AB})>0\;,
\ee
 if and only if $r\leq 1+d(d-1)/2$, and there exists a full rank matrix $X\in\mM_{d}(\mbb{C})$ 
 and a subspace $\mN_0\subset\mN$ with $\dim(\mN_0)=r-1$ such that
 \be\label{acon2}
 \supp(\rho^{AB})= X\mK\equiv\{XY\;\big|\;Y\in\mK\}\;,
 \ee
 where $\mK\equiv \{I\}\oplus\mN_0\equiv\{cI+N\;\big|\;c\in\mbb{C}\;;\;N\in\mN_0\}$.
}
 \begin{remark}
In the statement of Theorem~\ref{gcon} of the main text, we viewed $\supp(\rho^{AB})$ 
as a subspace of $\mbb{C}^d\otimes\mbb{C}^d$.
Here, for convenience of the proof, we use the isomorphism~\eqref{iso2} between 
$\mM_d(\mbb{C})$ and $\mbb{C}^d\otimes\mbb{C}^d$, and view the $\supp(\rho^{AB})$ 
as a subspace of $\mM_d(\mbb{C})$. The matrix $X$ in~\eqref{acon2} is related to 
the bipartite state $|x\ra^{AB}$ in~\eqref{con2} via~\eqref{iso2}. Recall also 
that a nilpotent matrix has a zero trace, and  is orthogonal to the identity 
matrix in the Hilbert-Schmidt inner product. This is consistent with the direct sum in the definition of $\mK$.
\end{remark}

\begin{proof}

Suppose there exist $X$ and $\mN_0$ as above such that $\supp(\rho^{AB})\subseteq X\mK$, and let
\be\label{w}
\rho^{AB}=\sum_{j=1}^{r}|w_j\lr w_j|\;,
\ee
where $|w_j\ra$ are the sub-normalized vectors given in~\eqref{w2}; i.e. 
$G(|w_j\ra)=0$ for $j>1$.
Since $W_j\in X\mK$, there exist constants $c_j\in\mbb{C}$ and matrices 
$Z_j\in\mN_{0}$ (corresponding to some sub-normalized vectors $|z_j\ra$) such that
\be\label{wj}
W_j=c_jX+XZ_j\;.
\ee
For $j>1$ we have
\be
0=G(|w_j\ra)=\left|\det(c_jX+XZ_j)\right|^{2/d}
=G(|x\ra)\left|\det(c_jI+Z_j)\right|^{2/d}
\ee
so that $\det(c_jI+Z_j)=0$ since $G(|x\ra)>0$. But since $Z_j$ is a nilpotent matrix
$\det(c_jI+Z_j)=(c_j)^d$. Hence, $c_j=0$ for $j>1$ and we denote by $c\equiv c_1$.
Note that $c\neq 0$ since otherwise we will get $G(\rho^{AB})=0$.

Therefore, the average G-concurrence of decomposition~\eqref{w} is given by
\be\label{firsteq}
\sum_{j=1}^{r}G(|w_j\ra)=G(|w_1\ra)=\left|\det(cX+XZ_1)\right|^{2/d}
=G(|x\ra)\left|c\right|^{2}\;.
\ee
Next, let
\be
\rho^{AB}=\sum_{k=1}^{m}|y_k\lr y_k|\;,
\ee
be another pure state decomposition of $\rho^{AB}$ with $m\geq r$ and 
$\{|y_k\ra\}$ some sub-normalized states. Then, there exists an 
$m\times r$ isometry $U=(u_{kj})$, i.e. $U^{\dag}U=I_r$, such that
\be
|y_k\ra=\sum_{j=1}^{r}u_{kj}|w_{j}\ra\;.
\ee
Using the form~\eqref{wj} with $c_1\equiv c$ and $c_2=c_3=...=c_r=0$, we get
\be
Y_k=\sum_{j=1}^{r}u_{kj}W_{j}=X\left(\sum_{j=1}^{r}u_{kj}\left(c\delta_{1j}I+Z_j\right)\right)
\equiv X\left(u_{k1}cI+N_k\right)\;,
\ee
where $N_k\equiv \sum_{j=1}^{r}u_{kj}Z_j\in\mN_{0}$ are nilpotent matrices. Consequently,
\be
G(|y_k\ra)=\left|\det\big(X\left(u_{k1}cI+N_k\right)\big)\right|^{2/d}=|u_{k1}|^2|c|^2G(|x\ra)\;,
\ee
where we used the fact that $\det(u_{k1}cI+N_k)=(u_{k1}c)^{d}$ since $N_k$ 
is nilpotent. Note that since the matrix $U$ is an isometry, $\sum_{k=1}^{m}|u_{k1}|^2=1$.
Therefore,
\be
\sum_{k=1}^{m}G(|y_k\ra)=|c|^2G\left(|x\ra\right)=\sum_{j=1}^{r}G(|w_j\ra)\;,
\ee
where the last equality follows from~\eqref{firsteq}. Therefore, all pure states 
decomposition of $\rho^{AB}$ have the same average G-concurrence so that $G(\rho^{AB})=G_a(\rho^{AB})$.

To prove the converse, suppose $G(\rho^{AB})=G_a(\rho^{AB})>0$, and consider the 
decomposition~\eqref{w2} of $\rho^{AB}$ as in the Lemma above.  
Consider the unnormalized state $\sigma^{AB}\equiv\rho^{AB}-|w_1\lr w_1|$.
The state $\sigma^{AB}=\sum_{j=2}^{r}|w_j\lr w_j|$ has zero $G$ concurrence 
since $G(|w_j\ra)=0$ for all $j=2,...,r$. Consider another decomposition of 
$\sigma^{AB}=\sum_{j=2}^r|y_j\lr y_j|$. Since $G(\rho^{AB})=G_a(\rho^{AB})$ 
we must have that $G(|y_j\ra)=0$ for all $j=2,...,r$. Otherwise, the 
decomposition $\rho^{AB}=|w_1\lr w_1|+\sum_{k=2}^{r}|y_k\lr y_k|$ will 
have a higher average G-concurrence than the decomposition in~\eqref{w2}. 
But since the $\{|y_j\ra\}$ decomposition was arbitrary, we conclude that 
all the states in the subspace $\mW\equiv\spa\{|w_2\ra,...,|w_r\ra\}$ 
have zero G-concurrence.

We now denote $W_1\equiv X$ and for $j=2,...,r$ we set $N_j\equiv X^{-1}W_j$. 
Note that with these notations,
$\supp(\rho^{AB})=\spa\{X,XN_2,XN_3,...,XN_r\}$. Since $X$ is invertible, 
and $\det(W)=0$ for any matrix $W$ in the span of $W_2,...,W_r$, we also 
have $\det(N)=0$ for any $N$ in the span of $N_2,...,N_r$. Let 
$N\in\spa\{N_2,...,N_r\}$ be a fixed matrix, and consider the two 
dimensional subspace $\mW_2\equiv\{X,XN\}\subset\supp(\rho^{AB})$.
We first show that $\mW_2$ does not contain a matrix with zero 
determinant that is linearly independent of $W\equiv XN$.
Indeed, if there exists a normalized matrix $Z\in\mW$ such that 
$\det(Z)=\det(W)=0$ with $W,Z$ being linearly independent, 
then $\mW_2=\spa\{W,Z\}$, and the rank 2 density matrix
\be
\sigma^{AB}=|w\lr w|+|z\lr z|
\ee
must have zero G-concurrence (recall that $G(|w\ra)=G(|z\ra)=0$). 
On the other hand, since $|x\ra\in\mW_2=\supp(\sigma^{AB})$, the 
density matrix $\sigma^{AB}$ must have a pure state decomposition 
containing $|x\ra$. Since $G(|x\ra)>0$ this decomposition does not 
have a zero average G-concurrence. Therefore, $0=G(\sigma^{AB})<G_a(\sigma^{AB})$. 
However, from Theorem~\ref{face} any density matrix $\sigma^{AB}$ 
with a support $\supp(\sigma^{AB})=\mW_2\subset\supp(\rho^{AB})$ 
has the same average G-concurrence for all pure state decompositions. 
We therefore get a contradiction with Theorem~\ref{face}, and thereby 
prove the assertion that $\mW_2$ does not contain another matrix 
with zero determinant that is linearly independent of $W$.

 Since $W=XN$ is the only matrix in $\mW_2$ with zero determinant 
 (up to multiplication by a constant), we must have that for any $\lambda\neq 0$
 \be
0\neq \det(X+\lambda W)=\det(X^{-1})\det(I+\lambda X^{-1}W)=\lambda^{d}\det(X^{-1})\det\left({1\over\lambda}I+ N\right)\;.
 \ee
Setting $t\equiv{1\over\lambda}$ we conclude that the polynomial
\be
f(t)\equiv\det\left(tI+ N\right)
\ee
is never zero for $t\neq 0$. On the other hand, $f(t)$ is a polynomial 
of degree $d$ and the coefficient of $t^d$ is one (note that it is the 
characteristic polynomial of $-N$). But since $t=0$ is the only root of $f(t)$, 
we must have $f(t)=t^d$; that is,
\be
\det\left(tI+ N\right)=t^n,\quad\forall\;t\in\mbb{C}\;.
\ee
Therefore, $N$ must be a \emph{nilpotent} matrix. Since $N$ was arbitrary, 
it follows that the subspace $\mN_0\equiv\spa\{N_2,...,N_r\}$ is a 
subspace of nilpotent matrices. This completes the proof.
\end{proof}

The G-concurrence is defined in~\eqref{defgcon} on pure bipartite states with the same local dimension.
For a pure bipartite state $|\psi\ra^{AB}\in\mH^A\otimes\mH^B$ 
with $\dim(\mH^A)<\dim(\mH^B)$ it is defined by:
\be
G(|\psi\ra^{AB})=\left(\det(\rho^A)\right)^{1/d_A}\;,
\ee
where $\rho^A\equiv\tr_B\left(|\psi\lr\psi|^{AB}\right)$ is the reduced 
density matrix, and $d_A\equiv\dim(\mH^A)$. With this extended definition, 
we have the following result:\\

\noindent\textbf{Corollary 8. }{\it
Let $|\psi^{ABC}\ra\in\mbb{C}^{d}\otimes\mbb{C}^{d}\otimes\mbb{C}^n$ be 
a pure tripartite state, with bipartite marginal $\rho^{AB}$. If
\be
G(\psi^{A|BC})=G(\rho^{AB})>0\;,
\ee
then $|\psi\ra^{ABC}=|\chi\ra^{AB}|\varphi\ra^{C}$ for some bipartite state
$|\chi\ra^{AB}\in\mbb{C}^{d}\otimes\mbb{C}^{d}$ and a vector $|\varphi\ra^C\in\mbb{C}^n$.
}

\begin{proof}
Since $|\psi\ra^{ABC}$ satisfies the disentangling condition, we get from 
Corollary~\ref{eoaeof} that $G(\rho^{AB})=G_a(\rho^{AB})>0$. Therefore, 
from the theorem above there exists a pure state decomposition of the 
marginal state $\rho^{AB}$ consisting of sub-normalized bipartite states
$\{|w_j\ra\}$ as in~\eqref{w2}, with the form
\begin{align}\label{aexp}
|w_1\ra&=\left(cX+XN_1\right)\otimes I^{B}|\phi_{+}\ra^{AB}\;,\nonumber\\
|w_j\ra&=XN_j\otimes I^{B}|\phi_{+}\ra^{AB}\;,\quad\forall\;j=2,...,r\;,
\end{align}
where $X$ is a full rank matrix, $N_j\in\mN_{0}$, and $c\in\mbb{C}$. 
Since all decompositions have the same average G-concurrence, we have
\be
G(\rho^{AB})=\sum_{j=1}^{r}G(|w_j\ra)=|\det(X)|^{2/d}c\;.
\ee
On the other hand, using the property that $\tr_B|\phi_+\lr\phi_+|^{AB}=I^A$, 
we get from~\eqref{aexp} that the marginal of $\rho^{AB}=\sum_{j=1}^{r}|w_j\lr w_j|^{AB}$ is given by:
\be
\rho^{A}=X(cI+N_1)(\bar{c}I+N_{1}^{\dag})X^{\dag}+X\sum_{j=2}^{r}N_jN_{j}^{\dag}X^{\dag}\;.
\ee
Therefore,
\be
G(|\psi^{A|BC}\ra)=\left(\det(\rho^A)\right)^{1/d}=|\det(X)|^{2/d}\left(\det\left(|cI+N_1|^2+\sum_{j=2}^{r}|N_j|^2\right)\right)^{1/d}\;,
\ee
where we used the notation $|A|\equiv\sqrt{AA^{\dag}}$ for any $d\times d$ matrix $A$.
Hence, the condition $G(|\psi^{A|BC}\ra)=G(\rho^{AB})=|\det(X)|^{2/d}|c|^2$ gives
\be\label{cz}
|c|^2=\left(\det\left(|cI+N_1|^2+\sum_{j=2}^{r}|N_j|^2\right)\right)^{1/d}\;.
\ee
Since $c\neq 0$ the matrix $cI+N_1$ is invertible. Denoting
\be
A\equiv |cI+N_1|\text{ and }B\equiv A^{-1}\left(\sum_{j=2}^{r}|N_j|^2\right) A^{-1}\;,
\ee
we get that~\eqref{cz} can be expressed as
\be
|c|^2=\left[\det\left(A(I+B)A\right)\right]^{1/d}=\left[\det\left(A^2\right)\right]^{1/d}\left[\det(I+B)\right]^{1/d}\;.
\ee
But since $N_1$ is nilpotent, we have
\be
\left[\det\left(A^2\right)\right]^{1/d}=\left[\det\left(|cI+N_1|\right)\right]^{2/d}=\left|\det(cI+N_1)\right|^{2/d}=|c|^2\;.
\ee
We therefore conclude that
\be\label{B}
\det(I+B)=1\;.
\ee
However, since $B\geq 0$, Eq.~\eqref{B} can hold only if $B=0$.
This in turn is possible only if $\sum_{j=2}^{r}|N_j|^2=0$; i.e. $N_j=0$ for all $j=2,...,d$. 
We therefore conclude that $\rho^{AB}$ is a pure state which implies that 
$|\psi\ra^{ABC}=|w_1\ra^{AB}|\varphi\ra^{C}$, where $|\varphi\ra^{C}$ is some pure state.
\end{proof}

Note that by combining the above corollary with Theorem~\ref{puremix} we get that 
the G-concurrence is fully monogamous (even for mixed tripartite states). 
This is to our knowledge the fist example of a monogamous measure of entanglement that is highly non-faithful.
 We conclude now with the 2-dimensional version of Theorem~\ref{gcon}. Recall 
 that the concurrence is the two dimensional G-concurrence.\\

\noindent\textbf{Corollary 9. }{\it
Let $\rho^{AB}$ be an entangled two qubit state with rank $r>1$, and let $E$ be 
any injective (up to local unitaries) measure of pure two qubit entanglement.
Then,
\begin{enumerate}
\item  If $E_f(\rho^{AB})=E_a(\rho^{AB})$ then $r=2$, and
$$
\rho^{AB}=\tr_C|W\lr W|^{ABC}
$$
is the 2-qubit marginal of the W-class state
\be\label{wstate}
|W\ra=\lambda_1|100\ra+\lambda_2|010\ra+\lambda_3|001\ra+\lambda_4|000\ra\;,
\ee
where $\lambda_j\in\mbb{C}$ and  $\sum_{j=1}^{4}|\lambda_j|^2=1$.
\item  Conversely, if $\rho^{AB}$ is a marginal of a state in the W-class then
$C_f(\rho^{AB})=C_a(\rho^{AB})$, where $C$ is the concurrence.
\end{enumerate}
}

\begin{proof}
\emph{Part 1:} In the two qubit case, we can also write $E(|\psi\ra^{AB})=g\left(C(|\psi\ra^{AB})\right)$, 
where $C(|\psi\ra^{AB})$ is the concurrence measure of entanglement, which is itself an injective 
measure. Moreover, since $E$ is injective so is the function $g$. Now, in~\cite{supWoo98}, 
Wootters showed that there always exists a pure state decomposition of $\rho^{AB}$ with each 
element in the decomposition being equal to $C_f(\rho^{AB})$. Denoting by $\{p_j,\;|\psi_j\ra^{AB}\}$ this decomposition, we have that
\be
E_f\left(\rho^{AB}\right)\leq \sum_jp_j g\left(C(|\psi_j\ra^{AB})\right)=g\left(C_f(\rho^{AB})\right)\;.
\ee
Similarly, there exists a pure state decomposition $\{q_k,\;|\phi_k\ra^{AB}\}$ of $\rho^{AB}$ such that for each $k$,
$C\left(|\phi_k\ra^{AB}\right)=C_a\left(\rho^{AB}\right)$~\cite{supGour2005pra}. Hence,
\be
E_a\left(\rho^{AB}\right)\geq \sum_kq_k g\left(C(|\phi_k\ra^{AB})\right)=g\left(C_a(\rho^{AB})\right)\;.
\ee
Therefore, the equality $E_f(\rho^{AB})=E_a(\rho^{AB})$ implies that $C_f(\rho^{AB})=C_a(\rho^{AB})$.
so that we can apply Theorem~\ref{gcon}. There is only one 1-dimensional subspace of $\mN$ 
(up to similarity), given by $\mN_0=\spa\left\{S\begin{pmatrix}0 & 1\\
0 & 0\end{pmatrix}S^{-1}\right\}$, where $S$ is a $2\times 2$ invertible matrix.
Now, from~\eqref{con2} together with the decomposition~\eqref{w2} we conclude 
that $\rho^{AB}$ can be expressed as:
\be
\rho^{AB}=|x\lr x|^{AB}+|y\lr y|^{AB}\;,
\ee
where $|x\ra^{AB}=X\otimes I^B|\phi_+\ra$ for some invertible matrix $X$, and
\be
|y\ra\equiv I\otimes S\begin{pmatrix}0 & 1\\
0 & 0\end{pmatrix}S^{-1}\big|x\ra^{AB}\;.
\ee
Therefore, $\rho^{AB}$ is a marginal of the tripartite state
\be
|\psi\ra^{ABC}=|x\ra^{AB}|0\ra^C+|y\ra^{AB}|1\ra^C\;.
\ee
Multiplying the above 3-qubit state by the SLOCC element $S^TX^{-1}\otimes S^{-1}\otimes I^C$, 
and using the symmetry $S^T\otimes S^{-1}|\phi_+\ra=|\phi_+\ra$ for any invertible matrix $S$, gives
\begin{align}
S^TX^{-1}\otimes S^{-1}\otimes I^C|\psi\ra^{ABC}&=\left(S^T\otimes S^{-1}|\phi_+\ra^{AB}\right)|0\ra^C+\left(S^T\otimes \begin{pmatrix}0 & 1\\
0 & 0\end{pmatrix}S|\phi_+\ra^{AB}\right)|1\ra^C\nonumber\\
&=|\phi_+\ra^{AB}|0\ra^C+\left(I^A\otimes \begin{pmatrix}0 & 1\\
0 & 0\end{pmatrix}|\phi_+\ra^{AB}\right)|1\ra^C\nonumber\\
&=|000\ra+|110\ra+|101\ra\;.
\end{align}
note that the state above is the W state after a local flip of the first qubit. Therefore, $|\psi\ra^{ABC}$ is in the $W$ class.

{\it Proof of Part 2:} We use for this part Wootters' formula for the concurrence.
Let $R=\sqrt{\rho^{1/2}\tilde{\rho}\rho^{1/2}}$ be Wootters matrix~\cite{supWoo98}, 
with $\rho\equiv\rho^{AB}$ and $\tilde{\rho}\equiv\sigma_y\otimes\sigma_y\rho^{*}\sigma_y\otimes\sigma_y$. 
Wootters showed in~\cite{supWoo98} that for entangled states 
$C_f(\rho^{AB})=\lambda_1-\lambda_2-\lambda_3-\lambda_4$, where
$\lambda_1,...,\lambda_4$ are the eigenvalues of $R$ in decreasing order. 
Furthermore, in~\cite{supLau03,supGour2005pra} it was shown that
$C_a(\rho^{AB})=\tr[R]=\lambda_1+\lambda_2+\lambda_3+\lambda_4$. 
Therefore, $C_f(\rho^{AB})=C_a(\rho^{AB})$ if and only if $R$ is a rank one matrix. 
A straightforward calculation shows that for the bipartite marginals, 
of the W-class states~\eqref{wstate}, $\tr[R^2]=\left(\tr[R]\right)^2$ 
so that the rank of $R$ is one (recall that $R$ is positive semidefinite). 
This completes the proof.
\end{proof}



\begin{thebibliography} {99}
		
				
		\bibitem{Coffman} V. Coffman, J. Kundu, and W. K. Wootters.
		Distributed entanglement.
		\textit{Phys. Rev. A}, 61:052306, 2000.
		\href{https://doi.org/10.1103/PhysRevA.61.052306}
		{doi:10.1103/PhysRevA.61.052306.}
		
		\bibitem{Horodecki2009} R. Horodecki, P. Horodecki, M. Horodecki, and K. Horodecki,
		Quantum entanglement.
		\textit{Rev. Mod. Phys.}, 81:865, 2009.
		\href{https://doi.org/10.1103/RevModPhys.81.865}
		{doi:10.1103/RevModPhys.81.865.}
		
		\bibitem{Koashi} M. Koashi and A. Winter.
		Monogamy of quantum entanglement and other correlations.
		\textit{Phys. Rev. A}, 69:022309, 2004.
		\href{https://doi.org/10.1103/PhysRevA.69.022309}
		{doi:10.1103/PhysRevA.69.022309.}
		
		
		\bibitem{Gour2005pra} G. Gour, D. A. Meyer, and B. C. Sanders.
		Deterministic entanglement of assistance and monogamy constraints.
		\textit{Phys. Rev. A}, 72:042329, 2005.
		\href{https://doi.org/10.1103/PhysRevA.72.042329}
		{doi:10.1103/PhysRevA.72.042329.}	
		
		\bibitem{Osborne} T. J. Osborne and F. Verstraete.
		General monogamy inequality for bipartite qubit entanglement.
		\textit{Phys. Rev. Lett.}, 96:220503, 2006.		
		\href{https://doi.org/10.1103/PhysRevLett.96.220503}
		{doi:10.1103/PhysRevLett.96.220503.}
		
		
		
		
		\bibitem{Ouyongcheng2007pra2} Y.-C. Ou and H. Fan,
		Monogamy inequality in terms of negativity for three-qubit states.
		\textit{Phys. Rev. A}, 75:062308, 2007.
		\href{https://doi.org/10.1103/PhysRevA.75.062308}
		{doi:10.1103/PhysRevA.75.062308.}
		
		
		
		\bibitem{Kim2009} J. S. Kim, A. Das, and B. C. Sanders.
		Entanglement monogamy of multipartite higher-dimensional quantum systems using convex-roof extended negativity.
		\textit{Phys. Rev. A}, 79:012329, 2009.
		\href{https://doi.org/10.1103/PhysRevA.79.012329}
		{doi:10.1103/PhysRevA.79.012329.}
		
		\bibitem{Zhuxuena2014pra} X. N. Zhu and S. M. Fei.
		Entanglement monogamy relations of qubit systems.
		\textit{Phys. Rev. A}, 90: 024304, 2014.
		\href{https://doi.org/10.1103/PhysRevA.90.024304}
		{doi:10.1103/PhysRevA.90.024304.}
		
		
		
		\bibitem{Bai} Y.-K. Bai, Y.-F. Xu, and Z. D. Wang.
		General monogamy relation for the entanglement of formation in multiqubit systems.
		\textit{Phys. Rev. Lett.}, 113:100503, 2014.
		\href{https://doi.org/10.1103/PhysRevLett.113.100503}
		{doi:10.1103/PhysRevLett.113.100503.}
		
		\bibitem{Choi} J. H. Choi and J. S. Kim.
		Negativity and strong monogamy of multiparty quantum entanglement beyond qubits.
		\textit{Phys. Rev. A}, 92:042307, 2015.
		\href{https://doi.org/10.1103/PhysRevA.92.042307}
		{doi:10.1103/PhysRevA.92.042307.}
		
		\bibitem{Luo} Y. Luo and Y. Li.
		Monogamy of $\alpha$th power entanglement measurement in qubit systems.
		\textit{Ann. Phys.}, 362:511-520, 2015.
		\href{https://doi.org/10.1016/j.aop.2015.08.022}
		{doi:10.1016/j.aop.2015.08.022.}
		
		
		\bibitem{Zhuxuena2015pra} X. N. Zhu and S. M. Fei.
		Entanglement monogamy relations of concurrence for $N$-qubit systems.
		\textit{Phys. Rev. A}, 92:062345, 2015.
		\href{https://doi.org/10.1103/PhysRevA.92.062345}
		{doi:10.1103/PhysRevA.92.062345.}
		
		\bibitem{Kumar} A. Kumar.
		Conditions for monogamy of quantum correlations in multipartite systems.
		\textit{Phys. Lett. A}, 380:3044-3050, 2016.
		\href{https://doi.org/10.1016/j.physleta.2016.07.032}
		{doi:10.1016/j.physleta.2016.07.032.}		
		
		\bibitem{Gour} G. Gour, S. Bandyopadhyay, and B. C. Sanders.
		Dual monogamy inequality for entanglement.
		\textit{J. Math. Phys.}, 48:012108, 2007.
		\href{https://doi.org/10.1063/1.2435088}
		{doi:10.1063/1.2435088.}
		
		\bibitem{Ouyongcheng} Y.-C. Ou.
		Violation of monogamy inequality for higher dimensional objects.
		\textit{Phys. Rev. A}, 75:034305, 2007.
		\href{https://doi.org/10.1103/PhysRevA.75.034305}
		{doi:10.1103/PhysRevA.75.034305.}
		
		\bibitem{Hiroshima2007prl} T. Hiroshima, G. Adesso and F. Illuminati.
		Monogamy inequality for distributed Gaussian entanglement.
		\textit{Phys. Rev. Lett.}, 98:050503, 2007.
		\href{https://doi.org/10.1103/PhysRevLett.98.050503}
		{doi:10.1103/PhysRevLett.98.050503.}
		
		\bibitem{Adesso} G. Adesso and F. Illuminati.
		Strong monogamy of bipartite and genuine multipartite entanglement: The Gaussian case.
		\textit{Phys. Rev. Lett.}, 99:150501, 2007.
		\href{https://doi.org/10.1103/PhysRevLett.99.150501}
		{doi:10.1103/PhysRevLett.99.150501.}
		
		\bibitem{Kim} J. S. Kim and B. C. Sanders.
		Generalized W-class state and its monogamy relation.
		\textit{J. Phys. A}, 41:495301, 2008.
		\href{https://doi.org/10.1088/1751-8113/41/49/495301}
		{doi:10.1088/1751-8113/41/49/495301.}
		
		\bibitem{Kim2010jpa} J. S. Kim and B. C. Sanders.
		Monogamy of multi-qubit entanglement using R\'{e}nyi entropy.
		\textit{J. Phys. A}, 43:445305, 2010.
		\href{https://doi.org/10.1088/1751-8113/43/44/445305}{doi:10.1088/1751-8113/43/44/445305.}
		
		\bibitem{Renxijun} X.-J. Ren and W. Jiang.
		Entanglement monogamy inequality in a $2\otimes 2\otimes 4$ system.
		\textit{Phys. Rev. A}, 81:024305, 2010.
		\href{https://doi.org/10.1103/PhysRevA.81.024305}
		{doi:10.1103/PhysRevA.81.024305.}
		
		\bibitem{Cornelio} M. F. Cornelio and M. C. de Oliveira.
		Strong superadditivity and monogamy of the R\'{e}nyi measure of entanglement.
		\textit{Phys. Rev. A}, 81:032332, 2010.
		\href{https://doi.org/10.1103/PhysRevA.81.032332}
		{doi:10.1103/PhysRevA.81.032332.}
		
		\bibitem{Streltsov} A. Streltsov, G. Adesso, M. Piani, and D. Bru{\ss}.
		Are general quantum correlations monogamous?
		\textit{Phys. Rev. Lett.}, 109:050503, 2012.
		\href{https://doi.org/10.1103/PhysRevLett.109.050503}
		{doi:10.1103/PhysRevLett.109.050503.}
		
		\bibitem{Cornelio2013pra} M. F. Cornelio.
		Multipartite monogamy of the concurrence.
		\textit{Phys. Rev. A}, 87:032330, 2013.
		\href{https://doi.org/10.1103/PhysRevA.87.032330}
		{doi:10.1103/PhysRevA.87.032330.}
		
		\bibitem{Liusiyuan} S.-Y. Liu, B. Li, W.-L. Yang, and H. Fan.
		Monogamy deficit for quantum correlations in a multipartite quantum system.
		\textit{Phys. Rev. A}, 87:062120, 2013.
		\href{https://doi.org/10.1103/PhysRevA.87.062120}
		{doi:10.1103/PhysRevA.87.062120.}
		
		\bibitem{Oliveira2014pra} T. R. de Oliveira, M. F. Cornelio, and F. F. Fanchini.
		Monogamy of entanglement of formation.
		\textit{Phys. Rev. A}, 89:034303, 2014.
		\href{https://doi.org/10.1103/PhysRevA.89.034303}
		{doi:10.1103/PhysRevA.89.034303.}
		
		\bibitem{Regula2014prl} B. Regula, S. D. Martino, S. Lee, and G. Adesso.
		Strong monogamy conjecture for multiqubit entanglement: the four-qubit case.
		\textit{Phys. Rev. Lett.}, 113:110501, 2014.
		\href{https://doi.org/10.1103/PhysRevLett.113.110501}
		{doi:10.1103/PhysRevLett.113.110501.}
		
		\bibitem{Salini} K. Salini, R. Prabhub, Aditi Sen(De), and Ujjwal Sen.
		Monotonically increasing functions of any quantum correlation can make all multiparty states monogamous.
		\textit{Ann. Phys.}, 348:297-305, 2014.
		\href{https://doi.org/10.1016/j.aop.2014.06.001}{doi:10.1016/j.aop.2014.06.001.}
		
		\bibitem{Hehuan} H. He and G. Vidal.
		Disentangling theorem and monogamy for entanglement negativity.
		\textit{Phys. Rev. A}, 91:012339, 2015.
		\href{https://doi.org/10.1103/PhysRevA.91.012339}
		{doi:10.1103/PhysRevA.91.012339.}
		
		\bibitem{Eltschka} C. Eltschka and J. Siewert.
		Monogamy equalities for qubit entanglement from Lorentz invariance.
		\textit{Phys. Rev. Lett.}, 114:140402, 2015.
		\href{https://doi.org/10.1103/PhysRevLett.114.140402}
		{doi:10.1103/PhysRevLett.114.140402.}
		
		\bibitem{Kumar2015} A. Kumar, R. Prabhu, A. Sen(de), and U. Sen.
		Effect of a large number of parties on the monogamy of quantum correlations.
		\textit{Phys. Rev. A}, 91:012341, 2015.
		\href{https://doi.org/10.1103/PhysRevA.91.012341}
		{doi:10.1103/PhysRevA.91.012341.}
		
		\bibitem{Lancien} Lancien \emph{et al}.
		Should entanglement measures be monogamous or faithful?
		\textit{Phys. Rev. Lett.}, 117:060501, 2016.
		\href{https://doi.org/10.1103/PhysRevLett.117.060501}
		{doi:10.1103/PhysRevLett.117.060501.}
		
		\bibitem{Lami} L. Lami, C. Hirche, G. Adesso, and A. Winter.
		Schur complement inequalities for covariance matrices and monogamy of quantum correlations.
		\textit{Phys. Rev. Lett.}, 117:220502, 2016.
		\href{https://doi.org/10.1103/PhysRevLett.117.220502}
		{doi:10.1103/PhysRevLett.117.220502.}
		
		\bibitem{Song} Song \emph{et al}.
		General monogamy relation of multiqubit systems in terms of squared R\'{e}nyi-$\alpha$ entanglement.
		\textit{Phys. Rev. A}, 93:022306, 2016.
		\href{https://doi.org/10.1103/PhysRevE.93.022306}
		{doi:10.1103/PhysRevE.93.022306.}
		
		\bibitem{Regula} B. Regula, A. Osterloh, and G. Adesso.
		Strong monogamy inequalities for four qubits.
		\textit{Phys. Rev. A}, 93:052338, 2016.
		\href{https://doi.org/10.1103/PhysRevA.93.052338}
		{doi:10.1103/PhysRevA.93.052338.}
		
		\bibitem{Luo2016pra} Y. Luo, T. Tian, L.-H. Shao, and Y. Li.
		General monogamy of Tsallis $q$-entropy entanglement in multiqubit systems.
		\textit{Phys. Rev. A}, 93: 062340, 2016.
		\href{https://doi.org/10.1103/PhysRevA.93.062340}
		{doi:10.1103/PhysRevA.93.062340.}
		
		\bibitem{Jung} E. Jung and D. Park.
		Testing the monogamy relations via rank-2 mixtures.
		\textit{Phys. Rev. A}, 94:042330, 2016.
		\href{https://doi.org/10.1103/PhysRevA.94.042330}
		{doi:10.1103/PhysRevA.94.042330.}
		
		
		\bibitem{Chengshuming} S. Cheng and M. J. W. Hall.
		Anisotropic invariance and the distribution of quantum correlations.
		\textit{Phys. Rev. Lett.}, 118:010401, 2017.
		\href{https://doi.org/10.1103/PhysRevLett.118.010401}
		{doi:10.1103/PhysRevLett.118.010401.}
		
		\bibitem{Allen}  G. W. Allen and D. A. Meyer.
		Polynomial monogamy relations for entanglement negativity.
		\textit{Phys. Rev. Lett.}, 118: 080402, 2017.
		\href{https://doi.org/10.1103/PhysRevLett.118.080402}
		{doi:10.1103/PhysRevLett.118.080402.}
		
		\bibitem{Liqiting} Q. Li, J. Cui, S. Wang, and G.-L. Long.
		Entanglement monogamy in three qutrit systems.
		\textit{Sci. Rep.}, 7:1946, 2017.
		\href{https://doi.org/10.1038/s41598-017-02066-8}
		{doi:10.1038/s41598-017-02066-8.}
		
		\bibitem{Camalet} S. Camalet.
		Monogamy Inequality for any local quantum resource and entanglement.
		\textit{Phys. Rev. Lett.}, 119: 110503, 2017.
		\href{https://doi.org/10.1103/PhysRevLett.119.110503}
		{doi:10.1103/PhysRevLett.119.110503.}
		
		\bibitem{Terhal} B. M. Terhal. 	
		IBM Journal of Research and Development,48(1):71-78, 2004.
		\href{https://doi.org/10.1147/rd.481.0071}{doi:10.1147/rd.481.0071.}
		
		\bibitem{Pawlowski} M. Pawlowski.
		Security proof for cryptographic protocols based only on the monogamy of Bell’s inequality violations.
		\textit{Phys. Rev. A}, 82:032313, 2010.
		\href{https://doi.org/10.1103/PhysRevA.82.032313}
		{doi:10.1103/PhysRevA.82.032313.}
		
		
		\bibitem{Gisin} N. Gisin, G. Ribordy, W. Tittel, and H. Zbinden.
		Quantum cryptography.
		\textit{Rev. Mod. Phys.}, 74:145, 2002.
		\href{https://doi.org/10.1103/RevModPhys.74.145}
		{doi:10.1103/RevModPhys.74.145.}
		
		\bibitem{Dur} W. D\"{u}r, G. Vidal, and J. I. Cirac.	
		Three qubits can be entangled in two inequivalent ways.
		\textit{Phys. Rev. A}, 62:062314, 2000.
		doi:10.1103/PhysRevA.62.062314.
		\href{https://doi.org/10.1103/PhysRevA.62.062314}
		{doi:10.1103/PhysRevA.62.062314.}
		
		
		\bibitem{Giorgi} G. L. Giorgi.
		Monogamy properties of quantum and classical correlations.
		\textit{Phys. Rev. A}, 84: 054301, 2011.
		\href{https://doi.org/10.1103/PhysRevA.84.054301}
		{doi:10.1103/PhysRevA.84.054301.}
		
		\bibitem{Prabhu} R. Prabhu, A. K. Pati, A. Sen(De), and U. Sen.
		Conditions for monogamy of quantum correlations: Greenberger-Horne-Zeilinger versus 
		W states.
		\textit{Phys. Rev. A}, 85:040102(R), 2012.
		\href{https://doi.org/10.1103/PhysRevA.85.040102}
		{doi:10.1103/PhysRevA.85.040102.}
		
		
		\bibitem{Ma} Ma \emph{et al}.
		Quantum simulation of the wavefunction to probe frustrated Heisenberg spin systems.
		\textit{Nat. Phys.}, 7:399, 2011.
		\href{https://doi.org/10.1038/nphys1919}{doi:10.1038/nphys1919.}
		
		
		\bibitem{Brandao} F. G. S. L. Brandao and A. W. Harrow, in \emph{Proceedings
			of the 45th Annual ACM Symposium on Theory of Computing,
			2013}.
		\href{http://dl.acm.org/citation.cfm?doid=2488608}
		{http://dl.acm.org/citation.cfm?doid=2488608.}
		
		\bibitem{Garcia} A. Garc\'{\i}a-S\'{a}ez and J. I. Latorre.
		Renormalization group contraction of tensor networks in three dimensions.
		\textit{Phys. Rev. B}, 87:085130, 2013.
		\href{https://doi.org/10.1103/PhysRevB.87.085130}
		{doi:10.1103/PhysRevB.87.085130.}
		
		
		\bibitem{Rao} Rao \emph{et al}.
		Multipartite quantum correlations reveal frustration in a quantum Ising spin system.
		\textit{Phys. Rev. A}, 88:022312, 2013.
		\href{https://doi.org/10.1103/PhysRevA.88.022312}
		{doi:10.1103/PhysRevA.88.022312.}
		
		\bibitem{Bennett} C. H. Bennett, in \emph{Proceedings of the FQXi 4th International
			Conference, Vieques Island, Puerto Rico, 2014},
		\href{http://fqxi.org/conference/talks/2014}
		{http://fqxi.org/conference/talks/2014.}
		
		\bibitem{Susskind} L. Susskind.
		Black hole complementarity and the Harlow-Hayden conjecture.
		\href{https://arxiv.org/abs/1301.4505} 
		{https://arxiv.org/abs/1301.4505.} 
		
		\bibitem{Lloyd} S. Lloyd and J. Preskill. 
		Unitarity of black hole evaporation in final-state projection models.
		\textit{J. High Energy Phys.}, 08:126, 2014.
		\href{https://doi.org/10.1007/JHEP08(2014)126}
		{doi:10.1007/JHEP08(2014)126.}
		
		\bibitem{Shor2001prl} P. W. Shor, J. A. Smolin, and B. M. Terhal.
		Nonadditivity of bipartite distillable entanglement follows from a conjecture on bound entangled Werner states. 
		\textit{Phys. Rev. Lett.}, 86:2681–2684, 2001.
		\href{https://doi.org/10.1103/PhysRevLett.86.2681}
		{doi:10.1103/PhysRevLett.86.2681.}
		
		\bibitem{Shor2004cmp} P. W. Shor.
		Equivalence of additivity questions in quantum information theory. 
		\textit{Commun. Math. Phys.}, 246(3):453-472, 2004.
		\href{https://doi.org/10.1007/s00220-003-0981-7}
		{doi:10.1007/s00220-003-0981-7.}
		
		\bibitem{Vollbrecht} K. G. H. Vollbrecht and R. F. Werner.
		Entanglement measures under symmetry.
		\textit{Phys. Rev. A}, 64:062307, 2001.
		\href{https://doi.org/10.1103/PhysRevA.64.062307}
		{doi:10.1103/PhysRevA.64.062307.} 
		
		
		\bibitem{gour2}
		G. Gour. 
		Family of concurrence monotones and its applications.
		\textit{Phys. Rev. A}, 71:012318, 2005.		
		\href{https://doi.org/10.1103/PhysRevA.71.012318}
		{doi:10.1103/PhysRevA.71.012318.}
		
		\bibitem{DiVincenzo} DiVincenzo \emph{et al}. 
		Entanglement of Assistance.
		\textit{Lecture Notes in Computer Science}, 1509:247, 1999.
		\href{https://doi.org/10.1007/3-540-49208-9\_21}
		{doi:10.1007/3-540-49208-9\_21.}
		
		\bibitem{HaydenJozaPetsWinter} P. Hayden, R. Jozsa, D. Petz, and A. Winter.
		Structure of states which satisfy strong subadditivity of quantum entropy with equality. 
		\textit{Commun. Math. Phys.}, 246(2):359-374, 2004.
		\href{https://doi.org/10.1007/s00220-004-1049-z}
		{doi:10.1007/s00220-004-1049-z.}
		
		
		\bibitem{Plenio2005} M. B. Plenio.
		Logarithmic negativity: a full entanglement monotone that is not convex.
		\textit{Phys. Rev. Lett.}, 95:090503, 2005.	
		\href{https://doi.org/10.1103/PhysRevLett.95.090503.}
		{doi:10.1103/PhysRevLett.95.090503.}
		Erratum \textit{Phys. Rev. Lett.}, 95:119902, 2005.
		\href{https://doi.org/10.1103/PhysRevLett.95.119902}
		{doi:10.1103/PhysRevLett.95.119902.}
		
		\bibitem{Vidal2000}  G. Vidal.
		Entanglement monotone.
		\textit{J. Mod. Opt.}, 47:355, 2000.
		\href{https://doi.org/10.1080/09500340008244048}
		{doi:10.1080/09500340008244048.}
		
		
		\bibitem{GourSpekkens}  G. Gour and R. W. Spekkens.
		Entanglement of assistance is not a bipartite measure nor a tripartite monotone. 
		\textit{Phys. Rev. A}, 73:062331, 2006.		
		\href{https://doi.org/10.1103/PhysRevA.73.062331}
		{doi:10.1103/PhysRevA.73.062331.}
		
		
		
		\bibitem{Gerstenhaber1958}  M. Gerstenhaber.
		On nilalgebras and linear varieties of nilpotent matrices (I). 
		\textit{Amer. J. Math.}, 80:614-622, 1958.		
		\href{https://doi.org/10.2307/2372773}{doi:10.2307/2372773.}		
		
		\bibitem{Woo98}
		W. K. Wootters.
		Entanglement of formation of an arbitrary state of two qubits.
		\textit{Phys. Rev. Lett.}, 80:2245, 1998.
		\href{https://doi.org/10.1103/PhysRevLett.80.2245}
		{doi:10.1103/PhysRevLett.80.2245.}
		
		
		
		
	\end{thebibliography}

\begin{thebibliography} {99}
	
\bibitem{supHehuan} H. He and G. Vidal.
Disentangling theorem and monogamy for entanglement negativity.
\textit{Phys. Rev. A}, 91:012339, 2015.
\href{https://doi.org/10.1103/PhysRevA.91.012339}
{doi:10.1103/PhysRevA.91.012339.}

\bibitem{supEltschka} C. Eltschka and J. Siewert.
Monogamy equalities for qubit entanglement from Lorentz invariance.
\textit{Phys. Rev. Lett.}, 114:140402, 2015.
\href{https://doi.org/10.1103/PhysRevLett.114.140402}
{doi:10.1103/PhysRevLett.114.140402.}

\bibitem{supAudenaert} K. M. R. Audenaert.
On a block matrix inequality quantifying the monogamy of the negativity of entanglement.
\textit{Lin. Multilin. Alg.}, 63(12):2526-2536, 2015.
\href{https://doi.org/10.1080/03081087.2015.1024193}
{doi/full/10.1080/03081087.2015.1024193.}

\bibitem{supLancien} Lancien \emph{et al}.
Should entanglement measures be monogamous or faithful?
\textit{Phys. Rev. Lett.}, 117:060501, 2016.
\href{https://doi.org/10.1103/PhysRevLett.117.060501}
{doi:10.1103/PhysRevLett.117.060501.}

\bibitem{supChengshuming} S. Cheng and M. J. W. Hall.
Anisotropic invariance and the distribution of quantum correlations.
\textit{Phys. Rev. Lett.}, 118:010401, 2017.
\href{https://doi.org/10.1103/PhysRevLett.118.010401}
{doi:10.1103/PhysRevLett.118.010401.}

\bibitem{supAllen}  G. W. Allen and D. A. Meyer.
Polynomial monogamy relations for entanglement negativity.
\textit{Phys. Rev. Lett.}, 118: 080402, 2017.
\href{https://doi.org/10.1103/PhysRevLett.118.080402}
{doi:10.1103/PhysRevLett.118.080402.}

\bibitem{supHillWotters} S. Hill and W. K. Wootters.
Entanglement of a pair of quantum bits. 
\textit{Phys. Rev. Lett.}, 78:5022, 1997.
\href{https://doi.org/10.1103/PhysRevLett.78.5022}
{doi:10.1103/PhysRevLett.78.5022.}

\bibitem{supKoashi} M. Koashi and A. Winter.
Monogamy of quantum entanglement and other correlations.
\textit{Phys. Rev. A}, 69:022309, 2004.
\href{https://doi.org/10.1103/PhysRevA.69.022309}
{doi:10.1103/PhysRevA.69.022309.}

\bibitem{supCoffman} V. Coffman, J. Kundu, and W. K. Wootters.
Distributed entanglement.
\textit{Phys. Rev. A}, 61:052306, 2000.
\href{https://doi.org/10.1103/PhysRevA.61.052306}
{doi:10.1103/PhysRevA.61.052306.}
\bibitem{supZhuxuena2014pra} X. N. Zhu and S. M. Fei,
Phys. Rev. A \textbf{90}, 024304 (2014).

\bibitem{supOsborne} T. J. Osborne and F. Verstraete.
General mnogamy inequality for bipartite qubit entanglement.
\textit{Phys. Rev. Lett.}, 96:220503, 2006.		
\href{https://doi.org/10.1103/PhysRevLett.96.220503}
{doi:10.1103/PhysRevLett.96.220503.}	

\bibitem{supOuyongcheng} Y.-C. Ou.
Violation of monogamy inequality for higher dimensional objects.
\textit{Phys. Rev. A}, 75:034305, 2007.
\href{https://doi.org/10.1103/PhysRevA.75.034305}
{doi:10.1103/PhysRevA.75.034305.}


\bibitem{supRenxijun} X.-J. Ren and W. Jiang.
Entanglement monogamy inequality in a $2\otimes 2\otimes 4$ system.
\textit{Phys. Rev. A}, 81:024305, 2010.
\href{https://doi.org/10.1103/PhysRevA.81.024305}
{doi:10.1103/PhysRevA.81.024305.}

\bibitem{supKim} J. S. Kim and B. C. Sanders.
Generalized W-class state and its monogamy relation.
\textit{J. Phys. A}, 41:495301, 2008.
\href{https://doi.org/10.1088/1751-8113/41/49/495301}
{doi:10.1088/1751-8113/41/49/495301.}

\bibitem{supOuyongcheng2007pra2} Y.-C. Ou and H. Fan,
Monogamy inequality in terms of negativity for three-qubit states.
\textit{Phys. Rev. A}, 75:062308, 2007.
\href{https://doi.org/10.1103/PhysRevA.75.062308}
{doi:10.1103/PhysRevA.75.062308.}

\bibitem{supLuo} Y. Luo and Y. Li.
Monogamy of $\alpha$th power entanglement measurement in qubit systems.
\textit{Ann. Phys.}, 362:511-520, 2015.
\href{https://doi.org/10.1016/j.aop.2015.08.022}
{doi:10.1016/j.aop.2015.08.022.}

\bibitem{supKim2009} J. S. Kim, A. Das, and B. C. Sanders.
Entanglement monogamy of multipartite higher-dimensional quantum systems using convex-roof extended negativity.
\textit{Phys. Rev. A}, 79:012329, 2009.
\href{https://doi.org/10.1103/PhysRevA.79.012329}
{doi:10.1103/PhysRevA.79.012329.}

\bibitem{supChoi} J. H. Choi and J. S. Kim.
Negativity and strong monogamy of multiparty quantum entanglement beyond qubits.
\textit{Phys. Rev. A}, 92:042307, 2015.
\href{https://doi.org/10.1103/PhysRevA.92.042307}
{doi:10.1103/PhysRevA.92.042307.}


\bibitem{supKumar} A. Kumar.
Conditions for monogamy of quantum correlations in multipartite systems.
\textit{Phys. Lett. A}, 380:3044-3050, 2016.
\href{https://doi.org/10.1016/j.physleta.2016.07.032}
{doi:10.1016/j.physleta.2016.07.032.}		

\bibitem{supBai}  Y.-K. Bai, Y.-F. Xu, and Z. D. Wang.
General monogamy relation for the entanglement of formation in multiqubit systems.
\textit{Phys. Rev. Lett.}, 113:100503, 2014.
\href{https://doi.org/10.1103/PhysRevLett.113.100503}
{doi:10.1103/PhysRevLett.113.100503.}

\bibitem{supOliveira2014pra} T. R. de Oliveira, M. F. Cornelio, and F. F. Fanchini.
Monogamy of entanglement of formation.
\textit{Phys. Rev. A}, 89:034303, 2014.
\href{https://doi.org/10.1103/PhysRevA.89.034303}
{doi:10.1103/PhysRevA.89.034303.}


\bibitem{supKim2016pra} J. S. Kim.
Tsallis entropy and general polygamy of multiparty quantum entanglement in arbitrary dimensions.
\textit{Phys. Rev. A}, 94:062338, 2016.
\href{https://doi.org/10.1103/PhysRevA.94.062338}
{doi:10.1103/PhysRevA.94.062338.}

\bibitem{supLuo2016pra} Y. Luo, T. Tian, L.-H. Shao, and Y. Li.
General monogamy of Tsallis $q$-entropy entanglement in multiqubit systems.
\textit{Phys. Rev. A}, 93: 062340, 2016.
\href{https://doi.org/10.1103/PhysRevA.93.062340}
{doi:10.1103/PhysRevA.93.062340.}

\bibitem{supKim2010jpa} J. S. Kim and B. C. Sanders.
Monogamy of multi-qubit entanglement using R\'{e}nyi entropy.
\textit{J. Phys. A}, 43:445305, 2010.
\href{https://doi.org/10.1088/1751-8113/43/44/445305}{doi:10.1088/1751-8113/43/44/445305.}

\bibitem{supCornelio} M. F. Cornelio and M. C. de Oliveira.
Strong superadditivity and monogamy of the R\'{e}nyi measure of entanglement.
\textit{Phys. Rev. A}, 81:032332, 2010.
\href{https://doi.org/10.1103/PhysRevA.81.032332}
{doi:10.1103/PhysRevA.81.032332.}

\bibitem{supSong} Song \emph{et al}.
General monogamy relation of multiqubit systems in terms of squared R\'{e}nyi-$\alpha$ entanglement.
\textit{Phys. Rev. A}, 93:022306, 2016.
\href{https://doi.org/10.1103/PhysRevE.93.022306}
{doi:10.1103/PhysRevE.93.022306.}

\bibitem{supGuo2013qip} Y. Guo, J. Hou, and Y. Wang.
Concurrence for infinite-dimensional quantum systems.
\textit{Quant. Inf. Process.}, 12:2641-2653, 2013.
\href{https://doi.org/10.1007/s11128-013-0552-6}
{doi:10.1007/s11128-013-0552-6.}


\bibitem{supGuo2013csb} Y. Guo and J. Hou.
Entanglement detection beyond the CCNR criterion for infinite-dimensions.
\textit{Chin. Sci. Bull.}, 58(11):1250-1255, 2013.
\href{https://doi.org/10.1007/s11434-013-5738-x}
{doi:10.1007/s11434-013-5738-x.}

\bibitem{supDonald1999pla} M. J. Donald and M. Horodecki.
Continuity of relative entropy of entanglement.
\textit{Phys. Lett. A}, 264:257, 1999.
\href{https://doi.org/10.1016/S0375-9601(99)00813-0}
{doi:10.1016/S0375-9601(99)00813-0.}

\bibitem{supGuo} The continuty of the convex roof extended entanglement measure
can be checked according to Proposition 2 in \cite{supGuo2013qip}, the continuty
of partial trace and partial transpose is proved in \cite{supGuo2013csb},
the continuty of the realtive entropy entanglement is proved in \cite{supDonald1999pla}.

\bibitem{supHaydenJozaPetsWinter} P. Hayden, R. Jozsa, D. Petz, and A. Winter.
Structure of states which satisfy strong subadditivity of quantum entropy with equality. 
\textit{Commun. Math. Phys.}, 246(2):359-374, 2004.
\href{https://doi.org/10.1007/s00220-004-1049-z}
{doi:10.1007/s00220-004-1049-z.}

\bibitem{supGourSpekkens} G. Gour and R. W. Spekkens.
Entanglement of assistance is not a bipartite measure nor a tripartite monotone. 
\textit{Phys. Rev. A}, 73:062331, 2006.		
\href{https://doi.org/10.1103/PhysRevA.73.062331}
{doi:10.1103/PhysRevA.73.062331.}

\bibitem{supUhlmann}  A. Uhlmann. Roofs and Convexity.
\textit{Entropy}, 12:1799-1832, 2010.
\href{https://doi.org/10.3390/e12071799}{doi:10.3390/e12071799.}

\bibitem{supGerstenhaber1958}  M. Gerstenhaber.
On nilalgebras and linear varieties of nilpotent matrices (I). 
\textit{Amer. J. Math.}, 80:614-622, 1958.		
\href{https://doi.org/10.2307/2372773}{doi:10.2307/2372773.}


\bibitem{supWoo98}
W. K. Wootters.
Entanglement of formation of an arbitrary state of two qubits.
\textit{Phys. Rev. Lett.}, 80:2245, 1998.
\href{https://doi.org/10.1103/PhysRevLett.80.2245}
{doi:10.1103/PhysRevLett.80.2245.}


\bibitem{supGour2005pra} G. Gour, D. A. Meyer, and B. C. Sanders.
Deterministic entanglement of assistance and monogamy constraints.
\textit{Phys. Rev. A}, 72:042329, 2005.
\href{https://doi.org/10.1103/PhysRevA.72.042329}
{doi:10.1103/PhysRevA.72.042329.}


\bibitem{supLau03} T. Laustsen, F. Verstraete, and S. J. van Enk. 
Local vs. joint measurements for the entanglement of assistance.
\textit{Quant. Inf. Comput}., 3:64, 2003.	
\href{https://arxiv.org/abs/quant-ph/0206192}{arXiv:0206192.}
	
	
\end{thebibliography}
\end{document}